\documentclass[aps,prd,superscriptaddress,footinbib,notitlepage,twocolumn]{revtex4-1}

\usepackage{tensor}
\usepackage{amsmath}
\usepackage{amsthm}
\usepackage{graphicx}
\usepackage{float}
\usepackage{comment}
\usepackage{amsfonts}
\usepackage{bm}
\usepackage{braket}
\usepackage{xcolor}
\usepackage[colorlinks=true,linkcolor=cyan,urlcolor=teal]{hyperref}

\usepackage{enumerate}
\usepackage{mathtools}
\usepackage{bbm}
\usepackage{mathtools}

\newcommand{\ii}{\mathrm{i}}
\newcommand{\dd}{\mathrm{d}}

\begin{document}

\title{Entanglement is better teleported than transmitted}
%\title{Quantum information is better teleported than transmitted}
%\title{Entanglement-harvesting-assisted communication is more efficient than direct communication through a quantum field}
\author{Koji Yamaguchi}
\affiliation{Department of Applied Mathematics, University of Waterloo, Waterloo, ON N2L 3G1, Canada}
\author{Achim Kempf}
\affiliation{Department of Applied Mathematics, University of Waterloo, Waterloo, ON N2L 3G1, Canada}
\affiliation{Department of Physics, University of Waterloo, Waterloo, ON N2L 3G1, Canada}
\affiliation{Institute for Quantum Computing, University of Waterloo, Waterloo, ON N2L 3G1, Canada}
%\affiliation{Perimeter Institute for Theoretical Physics, 31 Caroline St. N, Waterloo, Ontario, N2L 2Y5, Canada}

\begin{abstract}
We show that, for the purpose of quantum communication via a quantum field, it is essential to view the field not only as a medium for transmission but also as a source of entanglement that can aid in the communication task. To this end, we consider the quantum communication scenario where Alice is initially entangled with an ancilla and intends to communicate with Bob through a quantum field, so as to make Bob entangled with the ancilla. We find that if Alice and Bob communicate by directly coupling to the quantum field, then they can generate negativity between Bob and the ancilla only at orders that are higher than second perturbative order. We then present a protocol based on quantum teleportation in which Alice and Bob consume entanglement that they obtained from the field via interaction or harvesting. 
We show that this protocol can transfer negativity already to second perturbative order. 
%An analytic formula for the transferred negativity indicates that our protocol is optimal for transferring negativity in the case of a maximally entangled state and while using equal detectors for entanglement harvesting or generation. 

%The protocol in which Alice and Bob directly couples to the intermediary system, such as a field, can provide quantum channel capacity only from the fourth perturbative order and is highly sensitive to no-cloning constraints. Here, we show that protocols based on teleportation combined with entanglement harvesting or generation through the intermediary system, such as a field, can provide quantum channel capacity already to second perturbative order, while independent of to no-cloning constraints. 

%\textcolor{red}{abstract will be added here.}
\end{abstract}

\maketitle

\section{Introduction}
Conventional communication technologies focus on the transfer of classical information. For the development of  quantum communication technologies it is important to model not only the transfer of classical information but also of quantum information, such as the transfer of entanglement with an ancilla. 

Here, we investigate protocols for transferring entanglement with an ancilla, via an intermediary system such as a quantum field. Such protocols can be important, for example, for quantumly linking up smaller modules of a larger quantum memory or processor. 

Studies on communication through a quantum field have already revealed several new phenomena that have no analogs in classical communication. For example, it is known that classical information can be transmitted through a quantum field without transferring energy from a sender to a receiver \cite{jonsson_information_2015}. 
%In this protocol, the classical information is encoded in coherent superpositions of the ground state and the excited state of the sender. 
Since the receiver must provide energy to the field to extract the encoded information, this protocol is called quantum collect calling. Another protocol is quantum shockwave communication \cite{ahmadzadegan_quantum_2018}. A quantum shockwave can be formed by using senders that emit from spatially separated locations in spacetime, thereby mimicking a single sender on a superluminal trajectory. It was found that in this setup the multiple emitters can not only be used for conventional beam shaping (up to a classical shock wave pattern), but that it is possible to further shape the beam, beyond classical limitations, by preparing the emitters in a suitably entangled state. 

Of particular importance for our purposes here is the fact that quantum fields quantum fluctuate, even in the vacuum state. These quantum fluctuations represent a source of noise for any quantum system that interacts with the quantum field. The quantum fluctuations of a quantum field therefore generally hinder classical and quantum communication via a quantum field.  

In \cite{yamaguchi_superadditivity_2020}, a method was introduced by which a receiver of information from a quantum field can effectively reduce the impact of the quantum noise in the field. 
The method is based on the fact that the fluctuations of a quantum field at different spacetimes points are generally correlated \cite{summers_vacuum_1985,summers_bells_1987}.
%, which means that if the receiver uses multiple receiving devices at different spacetime points, these devices receive correlated noise. 
This means that a receiver that employs multiple receiving devices at different spacetime points is gaining some additional ability to tell noise from signal and can therefore effectively improve the signal to noise ratio. 

The receiver's devices can also be chosen spacelike separated from the sender. Those receiving devices receive no signal from the sender but they register quantum noise. This quantum noise is correlated with the noise in those receiving devices that also receive a signal. Therefore, the noise-only receivers are still able to help reduce the receiver's overall signal to noise ratio. In this sense, the classical channel capacity of communication through a quantum field can be superadditive, as shown in \cite{yamaguchi_superadditivity_2020}.

The phenomenon that correlated auxiliary noise can be used to improve the signal to noise ratio has been further explored in a study with neural networks \cite{ahmadzadegan_neural_2021}. These results, the so-called of ``Utilizing Correlated Auxiliary Noise" (UCAN) method, indicates opportunities for machine-learned quantum error correction. In this case, the correlated auxiliary noise consists of quantum fluctuations in environmental degrees of freedom that have inadvertently become entangled with the quantum processor. 
%This result suggests that a quantum field in communication plays a significant role as not only a carrier of signal but also a source of correlated fluctuation with which efficiency can be enhanced. 

For our purposes here, it is important to note that the occurrence of correlations between quantum field fluctuations at different points is closely related to the existence of entanglement in the vacuum state. For example, consider two localized quantum systems that couple to a field, i.e., so-called Unruh-DeWitt (UDW) detectors \cite{unruh_notes_1976,dewitt_quantum_1979}, or detectors for short. Trivially, if two detectors are coupling to the field at time-like or light-like separation they can become entangled through interaction via the field. However, even if the two detectors are coupling to the field in spatially separated regions, they can nevertheless become entangled, namely by extracting preexisting entanglement from the field. This phenomenon, called entanglement harvesting, has been extensively studied in flat and curved spacetimes, see, e.g., \cite{valentini_non-local_1991,reznik_entanglement_2003,reznik_violating_2005,retzker_detecting_2005,silman_long-range_2007,steeg_entangling_2009,olson_entanglement_2011,olson_extraction_2012,pozas-kerstjens_harvesting_2015,pozas-kerstjens_entanglement_2016,salton_acceleration-assisted_2015,nambu_entanglement_2013,sabin_extracting_2012,martin-martinez_sustainable_2013,lin_entanglement_2010,kukita_entanglement_2017,kukita_harvesting_2017,ng_unruh-dewitt_2018,henderson_harvesting_2018,henderson_entangling_2019,simidzija_harvesting_2018,braun_creation_2002,braun_entanglement_2005,hotta_duality_2020}. For example, the harvested entanglement can be used to detect spacetime curvatures and topologies \cite{steeg_entangling_2009,martin-martinez_spacetime_2016,saravani_spacetime_2016,kempf_replacing_2021,perche_geometry_2022}. Entanglement harvested from the field can be used not only to detect spacetime structures but also to improve the efficiency of communication. The quantum channel of communication through a quantum field using UDW detectors was introduced in \cite{cliche_relativistic_2010}. In \cite{koga_quantum_2018}, it has been shown that entanglement harvested or generated through a quantum field can be used to enhance the average teleportation fidelity. 
For an analysis of entanglement and transfer fidelity in a wider context, see also \cite{apollaro_entangled_2023}.

%More generally, since the complete information about the metric is contained in field commutators \cite{saravani_spacetime_2016}, the very notion of distance can be replaced by the notion of field correlation \cite{kempf_replacing_2021}, and the geometry of spacetime can be recovered by measuring field correlators with detectors \cite{perche_geometry_2022}. Entanglement harvested from the field can be used not only to detect spacetime structures but also to improve the efficiency of communication. The quantum channel of communication through a quantum field using UDW detectors was introduced in \cite{cliche_relativistic_2010}. In \cite{koga_quantum_2018}, it has been shown that entanglement harvested or generated through a quantum field can be used to enhance the average teleportation fidelity. For an analysis of entanglement and transfer fidelity in a wider context, see also \cite{apollaro_entangled_2023}.

In this paper, we study the ability of harvested entanglement to improve communication efficiency. We analyze two different scenarios (a) transmission and (b) teleportation. In the former setup, a sender and a receiver, Alice and Bob, communicate by simply coupling their detectors to the field. In the latter case, Alice and Bob first prepare entanglement through entanglement harvesting, and then they implement a teleportation protocol by consuming it. As a quantifier of communication, we adopt negativity with an ancillary system that is transferred from Alice to Bob. In perturbative analysis, we find that the negativity cannot be transmitted to the second order in the coupling constant, while the negativity can be teleported of the second perturbative order.

This paper is organized as follows. In Sec.~\ref{sec:setup}, we explain (a) transmission and (b) teleportation scenarios in more detail to clarify their difference. In Sec.~\ref{sec:direct_comm}, we prove a no-go theorem in case (a), stating that no negativity is transmitted to the second order of the coupling constant, regardless of the details of the interaction and the initial state. In Sec.~\ref{sec:EH-assisted}, we analyze case (b). In Sec.~\ref{sec:ordinary_QT}, we briefly review the ordinary teleportation protocol \cite{bennett_teleporting_1993}. In Sec.~\ref{sec:EH}, we introduce a setup for entanglement harvesting, which is commonly used in the literature. In Sec.~\ref{sec:QT_noisy_resource}, we propose a slightly adapted teleportation protocol.
%so that negativity is effectively transmitted by consuming the entanglement harvested or generated through a quantum field. 
We find that negativity can already be transferred to second perturbative order in this case.
%We prove an analytic formula for transferred negativity to the second order of the coupling constant. 
%In particular, we show that negativity can be transmitted to the second perturbative order and that our protocol is optimal when $\widetilde{A}A$ are initially in a maximally entangled state and identical detectors are used in entanglement harvesting or generation. Finally, the conclusions and outlook are presented in Sec.~\ref{sec:conclusion}. 
% analyze the efficiency of direct communication. We prove a no-go theorem, stating that no negativity can be transmitted to the second order of coupling constant, regardless of the detail of interaction between qubits and a field. In Sec.~\ref{sec:EH-assisted}, we first review the quantum teleportation protocol and entanglement harvesting. We then propose a slightly modified version of the quantum teleportation protocol, with which negativity can be transferred to the second order of coupling constant by consuming extracted entanglement by entanglement harvesting. In Sec.~\ref

Throughout this paper, we adopt natural units in which: $\hbar=c=1$.

%In \cite{saravani_spacetime_2016}, it was shown that the field correlators contain complete information about the metric and in \cite{kempf_replacing_2021} it was proposed to replace the very notion of distance by the notion of field correlation. This viewpoint has been further explored in \cite{perche_geometry_2022}, where it is demonstrated that the geometry of spacetime can be recovered by measuring field correlators with detectors. The fact that entanglement that has been harvested from a quantum field can be used to distinguish Minkowski spacetime from a Friedmann–Lema\^{i}tre–Robertson–Walker spacetime was already shown in \cite{steeg_entangling_2009}. 
%For the detection of topology from harvested entanglement, see \cite{martin-martinez_spacetime_2016}. Entanglement harvested from the field can be used not only to detect spacetime structures but also to improve the efficiency of communication. The quantum channel of communication through a quantum field using UDW detectors was introduced in \cite{cliche_relativistic_2010}. In \cite{koga_quantum_2018}, it has been shown that entanglement harvested or generated through a quantum field can be used to enhance the average teleportation fidelity. For an analysis of entanglement and transfer fidelity in a wider context, see also \cite{apollaro_entangled_2023}.

\section{Setup}\label{sec:setup}
In this paper, we investigate the efficiency of transmitting entanglement by communication through a quantum field. Concretely, we consider the following setup: A sender, Alice, has a qubit $A$, which is initially entangled with, and purified by an ancillary qubit $\widetilde{A}$. The ancillary qubit is assumed physically distant and will not take part in any interactions. Alice's task is to transmit the entanglement that her system, $A$, possesses with $\widetilde{A}$ to the qubit, $B$, of the receiver, Bob. 

The ability to communicate entanglement in this way could be very useful, for example, in order to create one large (error-corrected) quantum processor or memory from modules of smaller (error-corrected) quantum processors or memories. This is because a set of quantum modules only operates as one big unit, i.e., its Hilbert space dimension is the product of the Hilbert space dimensions of the modules, if entanglement can be spread at will across the modules so that all of this large Hilbert space is accessible. To this end, one needs to be able to communicate the entanglement that a qubit possesses with other qubits, i.e., with an ancilla, at will to qubits in other modules, for example, via the electromagnetic field.  
Hence, our setup can be applied: a qubit $A$ (in one module) is purified by a system $\widetilde{A}$ (that may be spread over one or multiple modules) and the goal is to communicate the entanglement that $A$ possesses with $\widetilde{A}$ to some qubit $B$ (in another module) via a quantum field such as a photon or a phonon field. 

\begin{figure}[htbp]
    \centering
    \includegraphics[width=5cm]{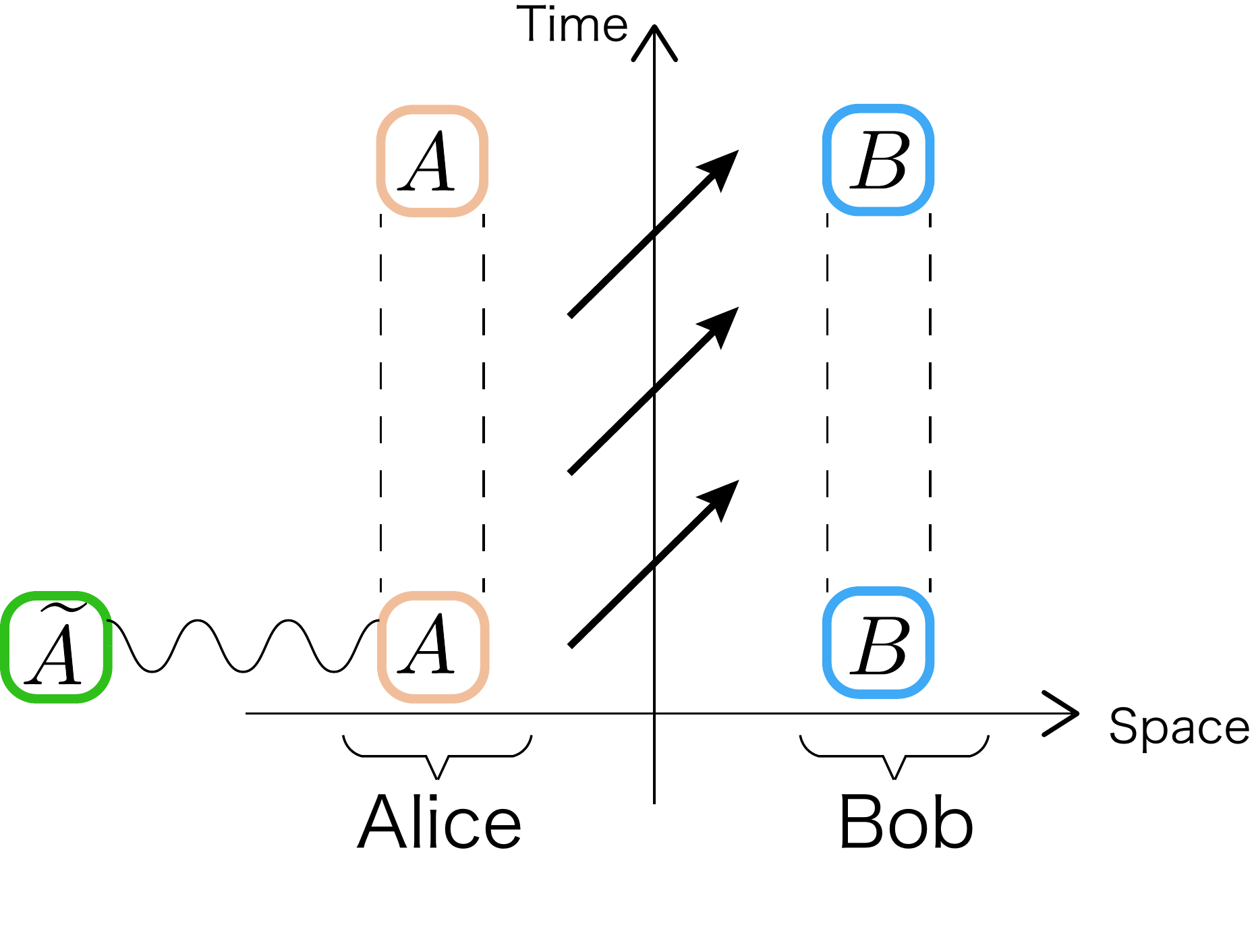}
    \caption{A schematic picture for transmission. A detector $A$ is initially purified by another detector $\widetilde{A}$. Quantum information of $A$ is encoded to the disturbance in the field by interaction, which then propagates through the spacetime and is later captured by another detector $B$. The arrows indicate the flow of disturbance carrying information. }
    \label{fig:direct}
\end{figure}

\begin{figure}[htbp]
    \centering
    \includegraphics[width=7.5cm]{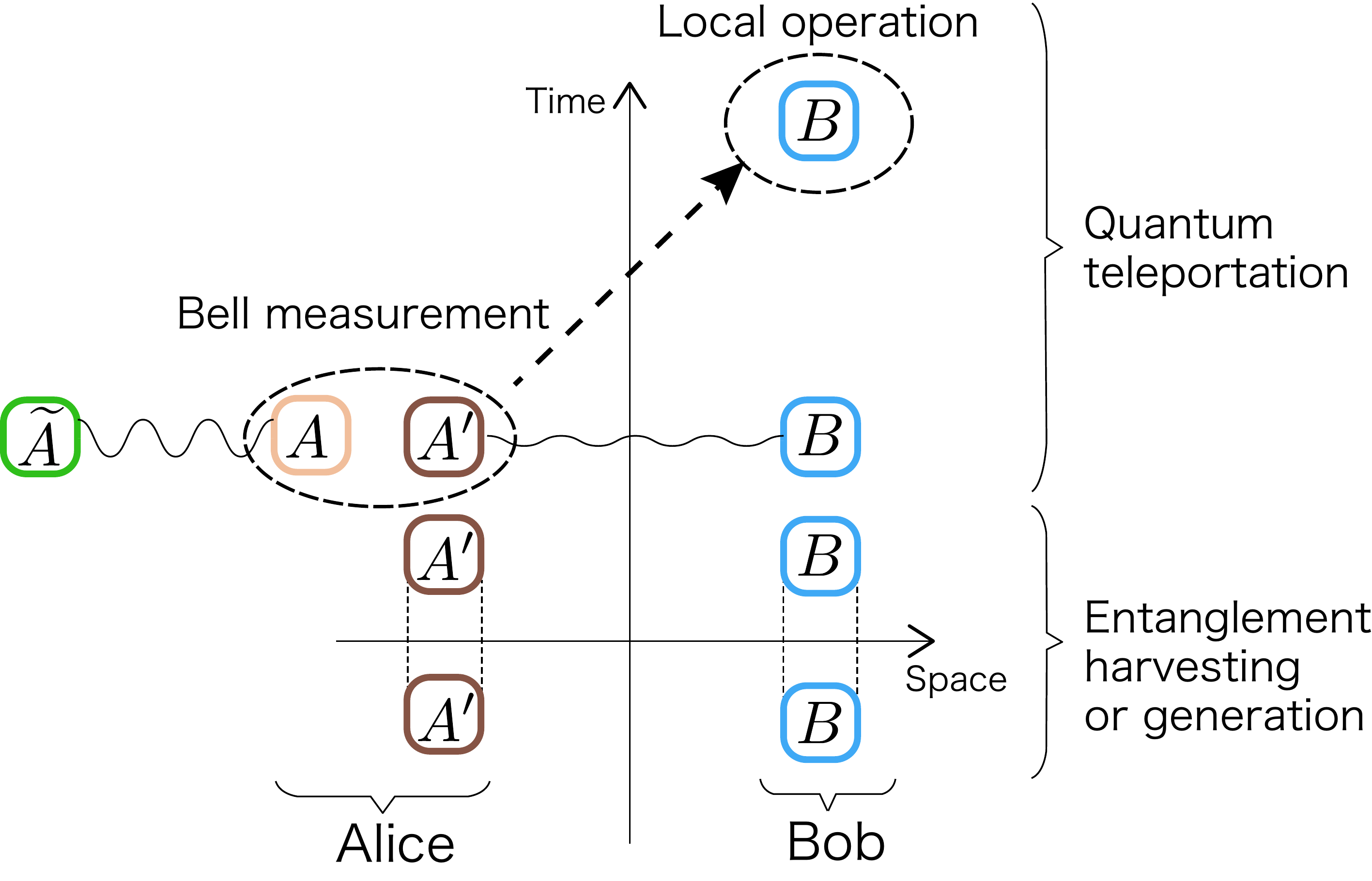}
    \caption{A schematic picture for teleportation assisted by entanglement generation or harvesting. Alice and Bob first make detectors $A'$ and $B$ entangled through interaction with the field. By consuming this entanglement, they perform a quantum teleportation protocol. The arrow with a dashed line represents the transmission of measurement outcome through classical wireless communication.}
    \label{fig:eh-assisted}
\end{figure}

We analyze two different scenarios: (a) Transmission: Alice simply couples her qubit $A$ to the field, and then Bob tries to pick up the encoded information from the field through the interaction between his qubit $B$ and the field. (b) Teleportation: Alice prepares another qubit $A'$. Alice and Bob first harvest or generate entanglement between qubits $A'$ and $B$ by interacting with the field. Concretely, we adopt UDW detector model, which is most commonly used in studies on entanglement harvesting. Alice and Bob perform a quantum teleportation protocol, thereby consuming the noisy entanglement resource generated or harvested through the field. Schematic images of these setups are shown in Figs.~\ref{fig:direct} and \ref{fig:eh-assisted}. 

%\textcolor{magenta}{comment on no-cloning constraint\cite{jonsson_transmitting_2018}? }

As a quantifier of the efficiency of communication, we adopt the negativity \cite{vidal_computable_2002} between $\widetilde{A}$ and $B$. Since negativity is an entanglement monotone, it does not increase by local operations and classical communication. Furthermore, negativity is a faithful entanglement measure for a bipartite qubit system in the sense that it vanishes if and only if the qubits are in a separable state \cite{horodecki_separability_1996}. 

We perturbatively calculate the negativity to the second order in the coupling constant in both setups (a) and (b). In case (a), we show that no negativity is transferred to the second order regardless of the details of the interaction Hamiltonian and the initial state. In case (b), we propose a teleportation protocol as a variant of the ordinary teleportation protocol \cite{bennett_teleporting_1993}. Although no closed formula is obtained in a general case, we find a condition under which the teleported negativity is given by a simple function of the Schmidt coefficient of the initial state of $\widetilde{A}A$ and the state of the entanglement resource $A'B$. 
In particular, when UDW detectors $A'$ and $B$ are identical and $\widetilde{A}A$ are initially maximally entangled, it is shown that the negativity between $\widetilde{A}$ and $B$ is precisely equal to the negativity between $A'$ and $B$ generated (or harvested) entanglement to the second order in the coupling constant. As a consequence, our protocol turns out to be optimal in this case. 
Based on these results, we find that quantum information, i.e., entanglement, is better teleported than transmitted in communication through a quantum field.

\section{No-go theorem for negativity transmission }\label{sec:direct_comm}
In this section, we analyze the efficiency of entanglement transfer in a transmission protocol. 
We show that no negativity is transferred to the second order in the coupling constant for generic interaction and initial states.

Let us first clarify the setup again. 
A sender, Alice, has a qubit $A$, which is initially purified by an auxiliary qubit $\widetilde{A}$. 
She encodes information of a qubit $A$ into an intermediary system $f$ through interaction between $A$ and $f$. The intermediary system is arbitrary, but we call it a quantum field for simplicity.
The disturbance in the quantum field caused by the qubit $A$ propagates through spacetime. 
A receiver, Bob, has a qubit $B$ coupled to the field, which picks up encoded information from the field. 

The initial state of the total system is assumed to be
\begin{align}
    \rho_{\widetilde{A}ABf}^{(0)}=\ket{\Psi}\bra{\Psi}_{\widetilde{A}A}\otimes \rho_B\otimes\rho_f,
\end{align}
where $\ket{\Psi}\bra{\Psi}_{\widetilde{A}A}$ denotes a pure state for qubits $\widetilde{A}A$, while $\rho_B$ and $\rho_f$ are arbitrary states for qubit $B$ and field $f$, respectively. We assume that the qubits $A$ and $B$ do not interact directly with each other and communicate only through interactions with the field. The general interaction Hamiltonian is given by
\begin{align}
    H_{\mathrm{I}}(t)=\sum_{i=A,B}H_i(t)=H_A(t)+H_B(t)
\end{align}
in the interaction picture. Here, $H_i(t)$ denotes the interaction Hamiltonian between the field and the qubit $i$ for $i=A,B$. 
We assume that each Hamiltonian $H_i$ is scaled by a coupling constant $\lambda_i$ and that both are of the same order, i.e., $\lambda_i=\mathcal{O}(\lambda)$ for $i=A,B$ for some parameter $\lambda>0$. 
We do not impose any further assumption on the interaction Hamiltonian. For example, a general form of $H_A(t)$ is given by
\begin{align}
    H_A(t)=\lambda_A\sum_{k}X^{(k)}_A\otimes \mathbb{I}_B\otimes O_f^{(k)}
\end{align}
for some Hermitian operators $X^{(k)}_A$ and $O_f^{(k)}$ on the qubit $A$ and the field $f$, respectively. Here, $\mathbb{I}_B$ denotes the identity operator. 

As a perturbative series with respect to $\lambda$, the time-evolution operator is given by the Dyson series:
\begin{align}
    U&=\mathcal{T}\exp\left(-i\int \dd t\, H_{\mathrm{I}}(t)\right)=\sum_{n=0}^\infty U^{(n)},
\end{align}
where
\begin{align}
    U^{(0)}&=\mathbb{I},\\ U^{(1)}&=(-\ii)\sum_{i=A,B}\int_{-\infty}^\infty \dd t\, H_{i}(t),\\
    U^{(2)}&=(-\ii)^2\sum_{i,j=A,B}\int_{-\infty}^\infty\dd t\int_{-\infty}^t\dd t'\, H_{i}(t)H_{j}(t')
\end{align}
to the second order in $\lambda$.
The reduced state for $\widetilde{A}B$ is expanded as
\begin{align}
    \rho_{\widetilde{A}B}=\rho_{\widetilde{A}B}^{(0)}+\rho_{\widetilde{A}B}^{(1)}+\rho_{\widetilde{A}B}^{(2)}+\mathcal{O}(\lambda^3),
\end{align}
where 
\begin{align}
    \rho_{\widetilde{A}B}^{(1)}&\coloneqq \mathrm{Tr}_{Af}\left(U^{(1)}\rho_{\widetilde{A}ABf}+\rho_{\widetilde{A}ABf}U^{(1)\dag}\right),\\
    \rho_{\widetilde{A}B}^{(2)}&\coloneqq \mathrm{Tr}_{Af}\left(U^{(2)}\rho_{\widetilde{A}ABf}+\rho_{\widetilde{A}ABf}U^{(2)\dag}+U^{(1)}\rho_{\widetilde{A}ABf}U^{(1)\dag}\right).
\end{align}
Similarly, the partial transpose of the quantum state for qubits $\widetilde{A}B$ is given by
\begin{align}
    \rho_{\widetilde{A}B}^{\top_{\widetilde{A}}}=\rho_{\widetilde{A}B}^{(0)\top_{\widetilde{A}}}+\rho_{\widetilde{A}B}^{(1)\top_{\widetilde{A}}}+\rho_{\widetilde{A}B}^{(2)\top_{\widetilde{A}}}+\mathcal{O}(\lambda^3),
\end{align}
where $\top_{\widetilde{A}}$ denotes the partial transpose with respect to the qubit system $\widetilde{A}$.

The negativity $\mathcal{N}(\rho_{\widetilde{A}B})$ of a bipartite system $\widetilde{A}B$ is defined as the sum of the absolute values of negative eigenvalues of the partial transposed matrix $\rho_{\widetilde{A}B}^{\top_{\widetilde{A}}}$. 
We now calculate the negativity perturbatively. 
Let 
\begin{align}
    \rho_{\widetilde{A}B}^{(0)\top_{\widetilde{A}}}=\sum_{p^{(0)}\in\sigma\left(\rho_{\widetilde{A}B}^{(0)\top_{\widetilde{A}}}\right)} p^{(0)}\, \Pi_{p^{(0)}}
\end{align}
be the eigenvalue decomposition, where $\sigma\left(\rho_{\widetilde{A}B}^{(0)\top_{\widetilde{A}}}\right)$ denotes the set of different eigenvalues of $\rho_{\widetilde{A}B}^{(0)\top_{\widetilde{A}}}$ and $\Pi_{p^{(0)}}$ is the projector on the eigenspace of $\rho_{\widetilde{A}B}^{(0)\top_{\widetilde{A}}}$ with an eigenvalue $p^{(0)}$.
When the detectors are coupled to the field, the eigenvalues of $\rho_{\widetilde{A}B}^{\top_{\widetilde{A}}}$, are perturbatively expanded as
\begin{align}
    p_i(\lambda)=p^{(0)}+\lambda p_i^{(1)}+\lambda^2p_i^{(2)}+\mathcal{O}(\lambda^3),
\end{align}
where $i$ is the label for the degeneracy of the eigenspace associated with the eigenvalue $p^{(0)}$. 
The negativity is defined as
\begin{align}
    \mathcal{N}\left(\rho_{\widetilde{A}B}\right)=\sum_{p_i\in\sigma\left(\rho_{\widetilde{A}B}^{\top_{\widetilde{A}}}\right),\,p_i<0}|p_i|. 
\end{align}
When the detectors do not interact with the field, i.e., $\lambda=0$, $\widetilde{A}$ and $B$ are not entangled. Therefore, all eigenvalues $p^{(0)}$ of $\rho_{\widetilde{A}B}^{(0)\top_{\widetilde{A}}}$ are non-negative. In this case, as is emphasized in Ref.~\cite{wen_transfer_2022}, $p_i\in\sigma\left(\rho_{\widetilde{A}B}^{\top_{\widetilde{A}}}\right)$ cannot become perturbatively negative unless $p^{(0)}=0$. Therefore, we will only analyze the perturbation of eigenvalues $p_i(\lambda)$ that vanishes if $\lambda=0$. 

We use the perturbation theory of eigenvalues, which is reviewed in Appendix~\ref{sec:appendix_perturbation_EV}. 
The first-order corrections $\lambda p_i^{(1)}$ for $p^{(0)}=0$ are given by the eigenvalues of the operator
\begin{align}
    \Pi_0\left(\rho_{\widetilde{A}B}^{(1)\top_{\widetilde{A}}}\right)\Pi_0
\end{align}
that is restricted to the eigenspace of $\rho_{\widetilde{A}B}^{(0)\top_{\widetilde{A}}}$ with eigenvalue $p^{(0)}=0$. Similarly, the second-order corrections $p_i^{(2)}$ are the eigenvalues of the following operator
\begin{align}
    &\Pi_0\rho_{\widetilde{A}B}^{(2)\top_{\widetilde{A}}}\Pi_0\nonumber\\
    &\quad -\Pi_0\rho_{\widetilde{A}B}^{(1)\top_{\widetilde{A}}}\sum_{p^{(0)}\in\sigma\left(\rho_{\widetilde{A}B}^{(0)\top_{\widetilde{A}}}\right)\setminus\{0\}}\frac{\Pi_{p^{(0)}}}{p^{(0)}}\rho_{\widetilde{A}B}^{(1)\top_{\widetilde{A}}}\Pi_0\label{eq:second_correction_operator}
\end{align}
in the eigenspace of $\rho_{\widetilde{A}B}^{(0)\top_A}$ with $p^{(0)}=0$.

There are two key facts that we use in the perturbative calculation. First, since $\rho_{\widetilde{A}B}^{(0)\top_{\widetilde{A}}}=\rho_{\widetilde{A}}^\top\otimes \rho_B$ and $\rho_{\widetilde{A}}^\top$ is invertible, we get
\begin{align}
    \Pi_0=\mathbb{I}_{\widetilde{A}}\otimes \pi_B,\label{eq:projector_zero}
\end{align}
where $\mathbb{I}_{\widetilde{A}}$ denotes the identity operator of system $\widetilde{A}$ and $\pi_B$ is the projector on the kernel of $\rho_B$. Second, the partial transpose operation $\top_{\widetilde{A}}$ commutes with the projector $\Pi_0=\mathbb{I}_A\otimes \pi_B$, i.e.,
\begin{align}
\begin{split}
    \Pi_0\left(O_{\widetilde{A}B}^{\top_{\widetilde{A}}}\right)&=\left(\Pi_0O_{\widetilde{A}B}\right)^{\top_{\widetilde{A}}}\\
    \left(O_{\widetilde{A}B}^{\top_{\widetilde{A}}}\right)\Pi_0&=\left(O_{\widetilde{A}B}\Pi_0\right)^{\top_{\widetilde{A}}}
\end{split}
    \label{eq:partial_transpose_formula}
\end{align}
for any linear operator $O_{\widetilde{A}B}$. See Appendix~\ref{sec:appendix_commute} for a proof. 

By using Eq.~\eqref{eq:partial_transpose_formula}, we have
\begin{align}
    &\Pi_0\left(\rho_{\widetilde{A}B}^{(1)\top_{\widetilde{A}}}\right)\Pi_0=\left(\Pi_0\rho_{\widetilde{A}B}^{(1)}\Pi_0\right)^{\top_{\widetilde{A}}}
\end{align}
Since $\pi_B\rho_B=\rho_B\pi_B=0$, we have
\begin{align}
    &\Pi_0\rho_{\widetilde{A}B}^{(1)}\Pi_0=0.
\end{align}
This means that all the eigenvalues of $\Pi_0\left(\rho_{\widetilde{A}B}^{(1)\top_{\widetilde{A}}}\right)\Pi_0$ vanish. Therefore, the negativity vanishes to the first order in the coupling constant.

Now, let us calculate the second-order corrections.
By using Eq.~\eqref{eq:partial_transpose_formula}, we have
\begin{align}
    &\Pi_0\rho_{\widetilde{A}B}^{(2)\top_{\widetilde{A}}}\Pi_0\nonumber\\
    &=\Pi_0\mathrm{Tr}_{Af}\left(\rho^{(2)\top_{\tilde{A}}}_{\tilde{A}ABf}\right)\Pi_0\nonumber\\
    &=\left(\Pi_0\mathrm{Tr}_{Af}\left(\rho^{(2)}_{\tilde{A}ABf}\right)\Pi_0\right)^{\top_{\tilde{A}}}\nonumber\\
    &=\left(\Pi_0\mathrm{Tr}_{Af}\left(\int_{-\infty}^\infty \dd t \int_{-\infty}^\infty \dd t' \, H_{B}(t)\rho_{\tilde{A}ABf}^{(0)}H_{B}(t')\right)\Pi_0\right)^{\top_{\tilde{A}}}.
\end{align}
In addition, we have
\begin{align}
    &\Pi_0\rho_{\widetilde{A}B}^{(1)\top_{\widetilde{A}}}\nonumber\\
    &=\Pi_0\mathrm{Tr}_{Af}\left(\rho^{(1)\top_{\tilde{A}}}_{\tilde{A}ABf}\right)\nonumber\\
    &=\mathrm{Tr}_{Af}\left(\mathbb{I}_{\widetilde{A}A}\otimes \pi_B\otimes \mathbb{I}_f\rho^{(1)\top_{\tilde{A}}}_{\tilde{A}ABf}\right)\nonumber\\
    &=(-\ii)\int_{-\infty}^\infty\mathrm{Tr}_{Af}\left(\mathbb{I}_{\widetilde{A}A}\otimes \pi_B\otimes \mathbb{I}_fH_B(t)\rho_{\widetilde{A}ABf}^{(0)\top_{\widetilde{A}}}\right)
\end{align}
and 
\begin{align}
    &\rho_{\widetilde{A}B}^{(1)\top_{\widetilde{A}}}\Pi_0\nonumber\\
    &=(-\ii)\int_{-\infty}^\infty\mathrm{Tr}_{Af}\left(\rho_{\widetilde{A}ABf}^{(0)\top_{\widetilde{A}}}H_B(t)\mathbb{I}_{\widetilde{A}A}\otimes \pi_B\otimes \mathbb{I}_f\right).
\end{align}
Therefore, the operator in Eq.~\eqref{eq:second_correction_operator} is independent of $H_A(t)$. This implies that the second-order corrections $p_{i}^{(2)}$ for $p^{(0)}=0$ are the same as those in the case where $H_A(t)=0$. If $H_A(t)=0$, the negativity between $\widetilde{A}$ and $B$ vanishes to all orders in $\lambda$. This implies that the negativity $\mathcal{N}\left(\rho_{\widetilde{A}B}\right)$ vanishes to the second order for a general interaction Hamiltonian in a transmission protocol.

So far, we have assumed that $\widetilde{A}$, $A$ and $B$ are qubits.  
The proof of this section can be directly extended to a general setup with any finite-dimensional quantum system instead of qubits as follows:
Without loss of generality, we can assume that the auxiliary system $\widetilde{A}$ that is added to purify the system $A$ is initially in an invertible state. Therefore, the projector $\Pi_0$ is given by Eq.~\eqref{eq:projector_zero}.
Following the same argument as in the above, we find that the negativity vanishes to the second order in the coupling constant. In this case, although the negativity is still an entanglement measure, it is not faithful in general since there are entangled states with vanishing negativity \cite{horodecki_separability_1996}.

%\textcolor{magenta}{Here we may compare our result with Robin's.}

\section{Entanglement transfer by entanglement-harvesting-assisted communication}\label{sec:EH-assisted}
In this section, we analyze a quantum teleportation protocol that makes use of the preexisting entanglement of the field. 
It is shown that by consuming the entanglement harvested from the field, negativity can be transferred to the second order in the coupling constant.

\subsection{Quantum state teleportation}\label{sec:ordinary_QT}
Let us first review the quantum teleportation protocol \cite{bennett_teleporting_1993}, which enables us to transmit quantum information perfectly by consuming a Bell state. 
As we shall see soon, correlations, including entanglement, with an ancillary system can also be transmitted by this protocol. 

Suppose that a sender, Alice, wants to transmit the state of a qubit $A$ to a receiver, Bob. We assume that this qubit $A$ is initially purified by $\widetilde{A}$. 
The quantum teleportation protocol consists of the following four steps:
\begin{enumerate}[(1)]
    \item First, Alice and Bob share qubits $A'$ and $B$ in a Bell state, given by
    \begin{align}
        \ket{\Phi_0}_{A'B}\coloneqq \frac{1}{\sqrt{2}}\left(\ket{g}_{A'}\ket{g}_B+\ket{e}_{A'}\ket{e}_B\right),
    \end{align}
    where $\ket{g}$ and $\ket{e}$ denote the ground and excited states, respectively.  
    Here, the qubit $A'$ and the qubit $B$ are accessible to Alice and Bob, respectively. 
    \item Alice performs a Bell measurement on $AA'$. Here, the Bell measurement is the projective measurement with respect to a Bell basis $\{\ket{\Phi_\mu}_{AA'}\}_{\mu=0}^3$, defined by 
    \begin{align}
        \ket{\Phi_\mu}_{AA'}\coloneqq\left(\sigma_\mu\otimes \mathbb{I}_{A'}\right)\frac{1}{\sqrt{2}}\left(\ket{\phi_g}_A\ket{g}_{A'}+\ket{\phi_e}_A\ket{e}_{A'}\right),\label{eq:bell_measurement_baisis}
    \end{align}
    where $\{\ket{\phi_g}_A,\ket{\phi_e}_A\}$ is an orthonormal basis for $A$. We here defined  $\sigma_0=\mathbb{I}$ and the Pauli operators $\sigma_1=\ket{\phi_e}\bra{\phi_g}+\ket{\phi_g}\bra{\phi_e}$, $\sigma_2=\ii(-\ket{\phi_e}\bra{\phi_g}+\ket{\phi_g}\bra{\phi_e})$ and $\sigma_3=\ket{\phi_e}\bra{\phi_e}-\ket{\phi_g}\bra{\phi_g}$.
    \item The measurement result $\mu=0,1,2,3$ is transmitted from Alice to Bob by classical communication.
    \item Bob performs a local operation $\sigma_\mu$ on his qubit $B$, depending on the outcome $\mu$. 
\end{enumerate}

Let us check that the state of the qubit $A$ is perfectly transferred to Bob in this protocol. 
Measurement operators are given by
\begin{align}
    M_\mu\coloneqq\mathbb{I}_{\widetilde{A}}\otimes \ket{\Phi_\mu}\bra{\Phi_\mu}_{AA'}\otimes \mathbb{I}_B.\label{eq:measurement_operator}
\end{align}
The initial state of $\widetilde{A}A$ is assumed to be an arbitrary pure state:
\begin{align}
    \ket{\Psi}_{\widetilde{A}A}=\sum_{i,j=e,g}c_{ij}\ket{i}_{\widetilde{A}}\ket{\phi_j}_{A},\quad c_{ij}\in\mathbb{C}.\label{eq:tele_initial_state}
\end{align}
After the measurement, the unnormalized selective post-measurement state is given by
\begin{align}
    M_{\mu} \left(\ket{\Psi}\bra{\Psi}_{\widetilde{A}A}\otimes\ket{\Phi_0}\bra{\Phi_0 
    }_{A'B} \right)M_\mu^\dag
\end{align}
Note that
\begin{align}
    &M_\mu \ket{\Psi}_{\widetilde{A}A}\otimes\ket{\Phi_0}_{A'B}\nonumber\\
    &=\left(\mathbb{I}_{\widetilde{A}}\otimes \ket{\Phi_\mu}\bra{\Phi_0}_{AA'}\otimes \mathbb{I}_B\right)\mathbb{I}_{\widetilde{A}A}\otimes\sigma_\mu^{\top} \otimes \mathbb{I}_{B}\ket{\Psi}_{\widetilde{A}A}\otimes\ket{\Phi_0}_{A'B}\nonumber\\
    &=\left(\mathbb{I}_{\widetilde{A}}\otimes \ket{\Phi_\mu}\bra{\Phi_0}_{AA'}\otimes \mathbb{I}_B\right)\mathbb{I}_{\widetilde{A}AA'}\otimes\sigma_\mu \ket{\Psi}_{\widetilde{A}A}\otimes\ket{\Phi_0}_{A'B}
\end{align}
holds, where we defined $\sigma_\mu$ on the subsystem $B$ with respect to the energy eigenbasis $\{\ket{g},\ket{e}\}$. 
%Here, we have used the fact that
%\begin{align}
%    O\otimes \mathbb{I}\ket{\Phi_0}=\mathbb{I}\otimes O^\top \ket{\Phi_0}
%\end{align}
%holds for any linear operator $O$, where $\top$ denotes the transpose operation with respect to the basis $\{\ket{g},\ket{e}\}$. 
The reduced state for $\widetilde{A}B$ is given by
\begin{align}
    &\mathrm{Tr}_{AA'}\left(M_{\mu} \left(\ket{\Psi}\bra{\Psi}_{\widetilde{A}A}\otimes\ket{\Phi_0}\bra{\Phi_0}_{A'B} \right)M_\mu^\dag\right)\nonumber\\
    &=\frac{1}{4}\left( \mathbb{I}_{\widetilde{A}}\otimes \sigma_\mu\right)\ket{\Psi}\bra{\Psi}_{\widetilde{A}B}\left( \mathbb{I}_{\widetilde{A}}\otimes \sigma_\mu\right)
\end{align}
if the measurement result is $\mu=0,1,2,3$. Here, $\ket{\Psi}_{\widetilde{A}B}$ is defined by
\begin{align}
    \ket{\Psi}_{\widetilde{A}B}=\sum_{i,j=e,g}c_{ij}\ket{i}_{\widetilde{A}}\ket{j}_{B}
\end{align}
with coefficients $c_{ij}$ defined in Eq.~\eqref{eq:tele_initial_state}. 
If Bob performs a local unitary operation $u_\mu$ on $B$ depending on the outcome $\mu$, the reduced state is given by
\begin{align}
   \rho_{\widetilde{A}B}= \frac{1}{4}\sum_\mu \left( \mathbb{I}_{\widetilde{A}}\otimes u_\mu\sigma_\mu\right)\ket{\Psi}\bra{\Psi}_{\widetilde{A}B}\left( \mathbb{I}_{\widetilde{A}}\otimes \sigma_\mu u_\mu^\dag\right).
\end{align}
For $u_\mu=\sigma_\mu$, the reduced state is given by
\begin{align}
    \rho_{\widetilde{A}B}=\ket{\Psi}\bra{\Psi}_{\widetilde{A}B},
\end{align}
implying that Bob recovers the state of the qubit $A$, including the correlation with $\widetilde{A}$.

\subsection{Entanglement harvesting and generation}\label{sec:EH}
In the ordinary quantum teleportation protocol, it is assumed that the sender and the receiver initially share a maximally entangled state. 
In this paper, we explore a way to enhance the efficiency of communication by entanglement generated or harvested through interaction with a quantum field. 

Here, we review a common setup of entanglement harvesting. Consider a scalar field $\phi$ in the $(d+1)$-dimensional Minkowski spacetime. 
If the mass of the field is $m$, the field operator is expanded as
\begin{align}
    &\phi(t,\bm{x})\nonumber\\
    &=\int\frac{\dd^d \bm{k}}{\sqrt{(2\pi)^d 2\omega_{\bm{k}}}}\left(a_{\bm{k}}^\dag e^{\ii (\omega_{\bm{k}}t-\bm{k}\cdot\bm{x})}+a_{\bm{k}} e^{-\ii (\omega_{\bm{k}}t-\bm{k}\cdot\bm{x})}\right),
\end{align}
where $\omega_{\bm{k}}\coloneqq \sqrt{|\bm{k}|^2+m^2}$. The creation and annihilation operators satisfy
\begin{align}
    \left[a_{\bm{k}},a_{\bm{k}'}^\dag\right]=\delta^{(d)}(\bm{k}-\bm{k}'),\quad \left[a_{\bm{k}},a_{\bm{k}'}\right]=\left[a_{\bm{k}}^\dag,a_{\bm{k}'}^\dag\right]=0.
\end{align}

Suppose that Alice and Bob have qubits $A'$ and $B$, respectively. These qubits play the role of an UDW detector, which is locally coupled to the field. For simplicity, we assume that the detectors are in inertial motion.
In the interaction picture, the interaction Hamiltonian is given by
\begin{align}
    H_{\mathrm{I}}(t)=\sum_{i=A',B}H_i(t),
\end{align}
where
\begin{align}
    H_{i}(t)\coloneqq \lambda_i\chi_i(t) \mu_i(t)\otimes \int \dd^d\bm{x}\,F_{i}(\bm{x}-\bm{x}_{i}(t))\phi(t,\bm{x}).
\end{align}
Here, $\lambda_i$ denotes the coupling constant, $\chi_i(t)$ is called the switching function that describes the temporal characterization of the interaction, and $F_{i}(\bm{x})$ is called the smearing function that characterizes the spatial extent of the detector $i=A',B$, where $\bm{x}_{i}(t)$ is the position of the center of mass of the detector. The monopole moment operator $\mu_i(t)$ is defined by
\begin{align}
    \mu_i(t)=\ket{e}\bra{g}_{i}e^{\ii \Omega_i t}+\ket{g}\bra{e}_i e^{-\ii \Omega_i t},
\end{align}
where $\Omega_i$ denotes the energy gap of the detector. 
Although the UDW-type interaction is simple, this model is known to be a good approximation of the light-matter interaction between atoms and the electromagnetic field when the exchange of angular momentum can be neglected \cite{pozas-kerstjens_entanglement_2016,martin-martinez_relativistic_2018}.
We assume that the coupling constants for the detectors are of the same order, parameterized by $\lambda$: $\lambda_i=\mathcal{O}(\lambda)$ for $i=A',B$. 

The time-evolution operator is expanded as
\begin{align}
    U&=\mathcal{T}\exp\left(-\ii\int\dd t H_{\mathrm{I}}(t)\right)\\
    &=\mathbb{I}+U^{(1)}+U^{(2)}+\mathcal{O}(\lambda^3)\\
    U^{(1)}&\coloneqq -\ii \int_{-\infty}^\infty\dd t H_{\mathrm{I}}(t)\\
    U^{(2)}&\coloneqq (-\ii)^2 \int_{-\infty}^\infty\dd t \int_{-\infty}^t \dd t' H_{\mathrm{I}}(t)H_{\mathrm{I}}(t').
\end{align}
We assume that the initial state is given by
\begin{align}
    \rho_{A'Bf}^{(0)}=\ket{g}\bra{g}_{A'}\otimes\ket{g}\bra{g}_B\otimes\rho_f.
\end{align}
The reduced state for the subsystems $A'B$ is given by
\begin{align}
    \rho_{A'B}&=\mathrm{Tr}_{f}\left(U\rho_{A'}\otimes\rho_B\otimes \rho_fU^\dag\right)\nonumber\\
    &=\rho_{A'B}^{(0)}+\rho_{A'B}^{(1)}+\rho_{A'B}^{(2)}+\mathcal{O}(\lambda^3)\\
    \rho_{A'B}^{(0)}&= \ket{g}\bra{g}_{A'}\otimes\ket{g}\bra{g}_B\\
    \rho_{A'B}^{(1)}&=\mathrm{Tr}_f\left(U^{(1)}\rho_{A'Bf}^{(0)}+\rho_{A'Bf}^{(0)}U^{(1)\dag}\right)\\
    \rho_{A'B}^{(2)}&=\mathrm{Tr}_f\left(U^{(2)}\rho_{A'Bf}^{(0)}+U^{(1)}\rho_{A'Bf}^{(0)}U^{(1)\dag}+\rho_{A'Bf}^{(0)}U^{(2)\dag}\right).\label{eq:second_order_rho}
\end{align}

For simplicity, we assume that the first moment of the field vanishes. For example, the vacuum state, squeezed states and thermal states with a quadratic Hamiltonian satisfy this condition. In this case, $\rho^{(1)}_{A'B}$ vanishes.

From Eq.~\eqref{eq:second_order_rho}, the matrix representation of $\rho_{A'B}$ in a basis $\{\ket{gg}_{A'B},\ket{ge}_{A'B},\ket{eg}_{A'B},\ket{ee}_{A'B}\}$ is given by
\begin{align}
    \rho_{A'B}=
    \begin{pmatrix}
    1-\mathcal{L}_{A'A'}-\mathcal{L}_{BB}&0&0&\mathcal{M}^*\\
    0&\mathcal{L}_{BB}&\mathcal{L}_{A'B}&0\\
    0&\mathcal{L}_{BA'}&\mathcal{L}_{A'A'}&0\\
    \mathcal{M}&0&0&0
    \end{pmatrix}+\mathcal{O}(\lambda^4)\label{eq:second_order_eh_state},
\end{align}
where we have defined
\begin{align}
    \mathcal{L}_{ij}&\coloneqq \lambda_i\lambda_j \int_{-\infty}^\infty\dd t \int_{-\infty}^\infty\dd t'\chi_i(t)\chi_j(t')\nonumber\\
    &\quad \times e^{-\ii(\Omega_i t-\Omega_j t')}W(t,\bm{x}_i;t',\bm{x}_j),\\
    \mathcal{M}&\coloneqq- \lambda_{A'}\lambda_B \int_{-\infty}^\infty \dd t \int_{-\infty}^t\dd t'\nonumber\\
    &\quad \left(e^{\ii( \Omega_{A'}t +\Omega_Bt')} \chi_{A'}(t)\chi_B(t')W(t,\bm{x}_{A'};t',\bm{x}_B)\right.\nonumber\\
    &\left.\quad +e^{\ii( \Omega_{B}t +\Omega_{A'}t')}\chi_{B}(t)\chi_{A'}(t')W(t,\bm{x}_{B};t',\bm{x}_{A'})\right),\\
    &W(t,\bm{x}_i;t',\bm{x}_j)\nonumber\\
    &\coloneqq\int\frac{\dd^d\bm{k}}{(2\pi)^d2\omega_{\bm{k}}}e^{-\ii \omega_{\bm{k}}(t-t')}e^{\ii\bm{k}\cdot(\bm{x}_i-\bm{x}_j)} |\widetilde{F}(\bm{k})|^2
\end{align}
and $ \widetilde{F}(\bm{k})\coloneqq \int\dd^d\bm{x}F(\bm{x})e^{\ii\bm{x}\cdot\bm{k}}$. 

Of the second order, there is a unique eigenvalue of $\rho_{A'B}^{\top_{A'}}$ that can be negative, given by
\begin{align}
    E\coloneqq\frac{1}{2}\left(\mathcal{L}_{A'A'}+\mathcal{L}_{BB}\right)-\frac{1}{2}\sqrt{(\mathcal{L}_{A'A'}-\mathcal{L}_{BB})^2+4|\mathcal{M}|^2}.
\end{align}
Therefore, we get
\begin{align}
    \mathcal{N}\left(\rho_{A'B}\right)&=\mathcal{N}^{(2)}\left(\rho_{A'B}\right)+\mathcal{O}(\lambda^3),\\ 
    \mathcal{N}^{(2)}\left(\rho_{A'B}\right)&\coloneqq \max\left\{0,-E\right\}.\label{eq:negativity_EH}
\end{align}

In particular, if $\mathcal{L}_{A'A'}=\mathcal{L}_{BB}$ holds, the formula is simplified and given by
\begin{align}
    \mathcal{N}^{(2)}\left(\rho_{A'B}\right)&= \max\left\{0,|\mathcal{M}|-\mathcal{L}_{A'A'}\right\}.\label{eq:negativity_simple_formula}
\end{align}
Intuitively, the condition $\mathcal{L}_{A'A'}=\mathcal{L}_{BB}$ means that $A'$ and $B$ are identical. 
The expression in Eq.~\eqref{eq:negativity_simple_formula} is commonly used in numerical calculations in the literature on entanglement harvesting. It has a simple physical interpretation: Since $\mathcal{L}_{A'A'}$ depends only on $\lambda_{A'}$, it is understood to be a contribution describing a local noise that $A'$ picks up. On the other hand, $\mathcal{M}$ depends on $\lambda_{A'}\lambda_B$ and is therefore related to a correlation between $A'B$ generated by the interaction. Equation~\eqref{eq:negativity_simple_formula} shows that the negativity becomes positive when the correlation term $\mathcal{M}$ is greater than the local noise term $\mathcal{L}_{A'A'}$. 

In the following subsection, we will use the quantum state in Eq.~\eqref{eq:second_order_eh_state} as a resource. 
We remark that this expression is valid independently of whether $A'$ and $B$ are space-like or causally separated.
We will not further explore the behavior of the negativity in Eq.~\eqref{eq:negativity_simple_formula} since this has already been analyzed in various setups in the literature.

\subsection{Quantum teleportation with noisy resource}\label{sec:QT_noisy_resource}

Here we calculate the amount of negativity transmitted by a quantum teleportation protocol by consuming an entangled resource state given by Eq.~\eqref{eq:second_order_eh_state}. 

We reviewed the ordinary teleportation protocol in Sec.~\ref{sec:ordinary_QT}, which perfectly transmits the entanglement by consuming a Bell pair.
Here, we propose a slightly adapted teleportation protocol since our resource is noisy.
Instead of Step (1),
\begin{enumerate}[(1')]
    \item Alice and Bob extract or generate entanglement from the field. As a consequence, they share a bipartite qubit system $A'B$ whose state is given by Eq.~\eqref{eq:second_order_eh_state}. 
\end{enumerate}
In addition, in our protocol, Bob performs a slightly adapted local operation in the last step: Instead of Step~(4), 
\begin{enumerate}[(4')]
    \item Bob performs a local unitary operation
    \begin{align}
     u_\mu=\sigma_\mu v_B,\label{eq:bob_local_op}
    \end{align}
    where
\begin{align}
    v_B\coloneqq e^{-\ii\varphi}\ket{e}\bra{e}_B+\ket{g}\bra{g}_B\label{eq:bob_phase_op}
\end{align}
and $\varphi\in\mathbb{R}$ is the argument of $\mathcal{M}$, i.e., $\mathcal{M}=|\mathcal{M}|e^{\ii \varphi}$.
\end{enumerate}

To perform the local operation in Step (4'), Bob has to know the parameter $\varphi$, i.e., the argument of $\mathcal{M}$, in addition to the measurement outcome $\mu$. However, this is not reducing the efficiency of our protocol. This is because $\mathcal{M}$ is the matrix element of the density matrix of the two-detector system right after the harvesting, and hence is fixed in every run of the setup if Alice and Bob agree beforehand on exactly how they will couple to the field. For comparison, we analyze the case where Bob does not know the value of $\varphi$ in Appendix~\ref{sec:phase_cancelling_op}. We find that the amount of teleported negativity is reduced if  he implements (4) instead of (4').

%We will calculate the teleported negativity under the assumption that the initial state for $A\widetilde{A}$ is given by a Bell state $\ket{\Psi_0}_{A\widetilde{A}}=(\ket{g}_A\ket{g}_{\widetilde{A}}+\ket{e}_A\ket{e}_{\widetilde{A}})/\sqrt{2}$. 
The amount of teleported negativity depends on the initial pure state $\ket{\Psi}_{\widetilde{A}A}$ of $\widetilde{A}A$. For example, the negativity between $\widetilde{A}B$ is upper bounded by the initial negativity between $\widetilde{A}$ and $A$ since the negativity is an entanglement monotone. To calculate the teleported negativity, we consider the Schmidt decomposition of the initial pure state $\ket{\Psi}_{\widetilde{A}A}$ of $\widetilde{A}A$, given by
\begin{align}
    \ket{\Psi}_{\widetilde{A}A}=\sum_{i=g,e}\sqrt{p_i}\ket{\psi_i}_{\widetilde{A}}\otimes\ket{\phi_i'}_{A},
\end{align}
where $\{\ket{\psi_i}\}_{i=g,e}$ and $\{\ket{\phi_i'}\}_{i=g,e}$ are orthonormal bases. It should be noted that in general, $\{\ket{\phi_i'}\}_{i=g,e}$ can be different from the basis $\{\ket{\phi_i}\}_{i=g,e}$ used to define the Bell measurement in Eq.~\eqref{eq:bell_measurement_baisis}.

After completing steps (1'), (2), (3), and (4'), the reduced state for $\widetilde{A}B$ is given by 
\begin{widetext}
\begin{align}
    \xi_{\widetilde{A}B}\coloneqq 2\sum_{i,j=g,e}\sqrt{p_ip_j}\ket{\psi_i}\bra{\psi_j}_{\widetilde{A}}\otimes \left(\sum_{k,l=g,e}\braket{\phi_k|\phi'_i}_{A}\braket{\phi_j'|\phi_l}_{A}\braket{k_{A'}|\eta_{A'B}|l_{A'}}\right),\label{eq:xi}
\end{align}
\end{widetext}
where we have defined
\begin{align}
    \eta_{A'B}&\coloneqq 
    \begin{pmatrix}
    \frac{1}{2}-\mathcal{L}&0&0&|\mathcal{M}|\\
    0&\mathcal{L}&\mathrm{Re}\left(\widetilde{\mathcal{L}_{A'B}}\right)&0\\
    0&\mathrm{Re}\left(\widetilde{\mathcal{L}_{A'B}}\right)&\mathcal{L}&0\\
    |\mathcal{M}|&0&0&\frac{1}{2}-\mathcal{L}
    \end{pmatrix}\nonumber\\
    &\quad +\mathcal{O}(\lambda^4),\label{eq:density_mat_noisy_qst}
\end{align}
$\mathcal{L}\coloneqq(\mathcal{L}_{A'A'}+\mathcal{L}_{BB})/2$ and $\widetilde{\mathcal{L}_{A'B}}\coloneqq \mathcal{L}_{A'B}e^{\ii\varphi}$.
See Appendix~\ref{sec:appendix_derivation_of_density_mat} for the derivation. 

For a general case, we do not have a closed formula for the negativity of $\eta_{\widetilde{A}B}$ since it is a complicated function depending on the Schmidt coefficients $\{p_i\}_i$, the inner products $\{\braket{\phi_i|\phi_k'}\}_{i,k}$ and the contributions $\mathcal{L}_{ij}$ and $|\mathcal{M}|$ that originate from the noisy entanglement resource state. 
For simplicity, we hereafter analyze the case where $\{\braket{\phi_i|\phi_k'}\}_{i,k}=\delta_{ik}$. 
This condition is satisfied when we adopt the basis $\{\ket{\phi_i'}\}_{i=g,e}$ that diagonalizes the initial state of $A$ as the basis $\{\ket{\phi_i'}\}_{i=g,e}$ in the definition of the Bell measurement. It should also be noted that this condition is satisfied when $\widetilde{A}A$ are initially maximally entangled since the initial state of $A$ is diagonalized in any orthonormal basis.  
In this case, we find that the negativity is given by
\begin{align}
    \mathcal{N}\left(\xi_{\widetilde{A}B}\right)&=\mathcal{N}^{(2)}\left(\xi_{\widetilde{A}B}\right)+\mathcal{O}(\lambda^3),\label{eq:negativity_teleportation_general}\\
    \mathcal{N}^{(2)}\left(\xi_{\widetilde{A}B}\right)&\coloneqq\max\{0,-E'\}\label{eq:negativity_teleportation_general_2nd},
\end{align}
where
\begin{align}
    E'&\coloneqq \mathcal{L}-\sqrt{\mathcal{L}^2(1-4p(1-p))+4p(1-p)|\mathcal{M}|^2}.\label{eq:negative_ev}
\end{align}
Here, we defined $p\coloneqq p_g$, which implies $p_e=1-p$. 

Let us now check the consistency of our formula Eq.~\eqref{eq:negativity_teleportation_general} with the monotonicity of the negativity.
We first remark that 
\begin{align}
    \mathcal{N}\left(\xi_{\widetilde{A}B}\right)\leq \mathcal{N}\left(\ket{\Psi}\bra{\Psi}_{\widetilde{A}A}\right)
\end{align}
holds for $\lambda\ll 1$ since the left hand side is of the second order in $\lambda$, while the right hand side is independent of $\lambda$. 

To prove
\begin{align}
    \mathcal{N}\left(\xi_{\widetilde{A}B}\right)\leq \mathcal{N}\left(\rho_{A'B}\right),\label{eq:monotonicity_of_negativity}
\end{align}
we consider two different cases: (1) When $|\mathcal{M}|^2< \mathcal{L}^2$, 
\begin{align}
    -E'\leq \sqrt{\mathcal{L}^2}-\mathcal{L}=0,
\end{align}
where the equality holds for $p=0$. Therefore, $\mathcal{N}^{(2)}\left(\xi_{\widetilde{A}B}\right)=0$ and hence Eq.~\eqref{eq:monotonicity_of_negativity} holds. (2) When $|\mathcal{M}|^2\geq \mathcal{L}^2$, we have
\begin{align}
    -E'\leq \sqrt{|\mathcal{M}|^2}-\mathcal{L}\leq -E,
\end{align}
where the equality in the first inequality holds for $p=1/2$.
Therefore, Eq.~\eqref{eq:monotonicity_of_negativity} holds.

When $\widetilde{A}$ and $A$ are initially in a maximally entangled state, i.e., $p=1/2$, the formula in Eq.~\eqref{eq:negativity_teleportation_general_2nd} is simplified and given by
\begin{align}
    \mathcal{N}^{(2)}\left(\xi_{\widetilde{A}B}\right)= \max\left\{0,|\mathcal{M}|-\mathcal{L}\right\}.
\end{align}
If the detectors $A'$ and $B$ are identical, the formula is further simplified and given by
\begin{align}
    \mathcal{N}^{(2)}\left(\xi_ {\widetilde{A}B}\right)&= \max\left\{0,|\mathcal{M}|-\mathcal{L}_{A'A'}\right\},
\end{align}
implying that
 \begin{align}
    \mathcal{N}^{(2)}\left(\xi_{\widetilde{A}B}\right)= \mathcal{N}^{(2)}\left(\rho_{A'B}\right).\label{eq:optimality}
\end{align}
Therefore, in this case, our protocol is an optimal way to transmit the entanglement by consuming the harvested entanglement. Furthermore, it also shows that the negativity extracted from the field in entanglement harvesting is always useful in teleporting the entanglement to the second order in the coupling constant.

It should be noted that our formula in Eq.~\eqref{eq:negativity_teleportation_general} is valid as long as the reduced state of the UDW detectors $A'B$ is given by Eq.~\eqref{eq:second_order_eh_state} for some $\mathcal{L}_{ij}$ and $\mathcal{M}$. In other words, our formula is independent of the details of $\mathcal{L}_{ij}$ and $\mathcal{M}$. Therefore, our result here is applicable to more general setups, e.g., a teleportation protocol using entanglement in detectors that is generated or harvested through interaction with a quantum field in a curved spacetime.

\section{Conclusions and outlook}\label{sec:conclusion}
In this paper, we investigated the efficiency of entanglement transfer through an intermediate system, such as a quantum field. Concretely, we considered the following setup: a sender, Alice, possesses a qubit $A$, which is initially entangled with an auxiliary system $\widetilde{A}$, while a receiver, Bob, possesses a qubit $B$. The aim is to transfer the entanglement that $A$ initially possesses with $\tilde{A}$ to $B$ so that then $B$ possesses this entanglement with $\tilde{A}$. To this end, Alice and Bob are allowed to each interact with an intermediary system such as a quantum field. We compared two different strategies that Alice and Bob may adopt, namely (a) transmission and (b) teleportation. In case (a), Alice and Bob only couple their qubits to the field. We proved that, regardless of the details of the interaction between the qubits and the intermediary system, in this way no negativity can be generated between $\widetilde{A}$ and $B$ to the second order in the coupling constant. 
In case (b), Alice and Bob first each use an UDW-type interaction to generate or harvest entanglement via the quantum field. For this purpose, Alice is using an additional qubit $A'$, rather than her qubit $A$. Crucially, the negativity between the qubits $A',B$ does become positive already to the second order in the coupling constant. Alice and Bob then use a slightly adapted version of the quantum teleportation protocol which consumes this noisy resource of entanglement in order to teleport some of Alice entanglement with the ancilla to Bob.  

For the case where the initial state of $A$ is diagonalized in a basis that is used to define a Bell measurement, we were able to derive a simple analytic formula for the amount of transferred negativity, given by Eq.~\eqref{eq:negativity_teleportation_general_2nd}. The formula shows that when the detectors $A'$ and $B$ are identical and $\widetilde{A}A$ are initially maximally entangled, then our protocol is optimal in the sense that it transfers exactly the same amount of negativity from the $\tilde{A}A$ system to the $\tilde{A}B$ system as the amount of negativity that Alice and Bob's $A'B$ system had obtained by interacting with the field. 

In summary, if Alice tries to transmit some of her initial entanglement with an ancilla $\tilde{A}$ to Bob by means of Alice and Bob merely coupling to an intermediate quantum field, then this can only succeed to higher than second order in the coupling constant, irrespective of how they each interact with the quantum field. Our finding is that, nevertheless, it is possible for Alice to transfer some of her entanglement to Bob through the quantum field already to second order in the coupling constant. To this end, Alice and Bob can use the fact that they become partially entangled when interacting with the quantum field. They  consume this entanglement to teleport some of Alice's entanglement with the ancilla to Bob, which can be done already to second order. This result is consistent with the result in \cite{koga_quantum_2018}, where it is shown that the average teleportation fidelity is increased by consuming the harvested entanglement. 

The above results can be extended to the scenario in which Alice  distributes her entanglement with the ancilla $\tilde{A}$ to multiple Bobs, $B_i$, for $i=1,2,\cdots,N$. First, we notice that no negativity can be transmitted to an $\widetilde{A}B_i$ system to second order in the coupling constant. This is because the corresponding calculation for a single Bob, in Sec.~\ref{sec:direct_comm}, still applies. Second, the teleportation-based protocol still applies and can transfer entanglement to second order from the $\tilde{A}A$ system to the $\tilde{A}B_i$ systems. Concretely, let us assume that the $N$ receiving devices are located in space-like separated regions.  
After the entanglement harvesting, the reduced states of each of the bi-partite systems $A'B_i$ takes the form of Eq.~\eqref{eq:second_order_eh_state}, though the actual matrix elements will differ for each $i$, except if the Bobs are distributed symmetrically around Alice. After Alice performs the Bell measurement, she classically transmits the outcome to the $N$ Bobs. Each Bob then performs the corresponding local unitary operation defined in Eq.~\eqref{eq:bob_local_op}, and thereby obtains an amount of negativity with $\tilde{A}$ given by Eq.~\eqref{eq:negativity_teleportation_general_2nd}. It  should be very interesting to combine this result with the entanglement monogamy constraint to derive non-perturbative absolute upper bounds on the amount of entanglement that can be harvested from quantum fields, as a function of the type of field and the type of interaction between the field and localized probes such as UDW detectors.

%This implies that Alice can simultaneously establish bipartite entanglement with each Bob by using the teleportation protocol. To perform the local operation in Eq.~\eqref{eq:bob_local_op}, $i$th Bob has to know the phase $\varphi$ of the matrix element $\mathcal{M}$ of the reduced state for $A'B_i$. Thus, different Bobs may need to implement different unitary operations if the designs of the detectors differ from each other. If Bobs are ignorant of the phase $\varphi$, the transmitted negativity can be degraded, as shown by the arguments in Appendix~\ref{sec:phase_cancelling_op}. 

%It should be interesting to study transmission and teleporation protocols for entanglement transfer in a more general setup where multiple senders and receivers collaborate to transfer entanglement. Such a case is a generalization of a classical multiple-input-multiple-output (MIMO) system, which may be called a quantum MIMO (QMIMO) system. \textcolor{magenta}{Benefits:}
%Exploring a way to improve the efficiency is essential to connect modules in a quantum computer and built totally integrated quantum networks. 

Our results in this paper show that, for the purpose of transferring entanglement with an ancilla through a quantum field, the direct method of transmission by simply coupling the sender and receiver to the field, is not optimal. We found that it is more efficient for Alice and Bob first to use the field to either generate entanglement or to harvest preexisting  entanglement and then to use quantum teleportation while consuming this entanglement. 

This indicates that, for the purpose of quantum communication through a quantum field, it is essential to view the  quantum field not only as a medium of propagation but also as a means to generate or harvest entanglement that can serve as a resource for the quantum communication.

\begin{acknowledgments}
KY acknowledges support from the JSPS Overseas Research Fellowships. AK acknowledges support through a Discovery Grant of the National Science and Engineering Council of Canada (NSERC) and a Discovery Project grant of the Australian Research Council (ARC). 
\end{acknowledgments}

\bibliographystyle{apsrev4-1}
\bibliography{references}

\begin{thebibliography}{46}%
\makeatletter
\providecommand \@ifxundefined [1]{%
 \@ifx{#1\undefined}
}%
\providecommand \@ifnum [1]{%
 \ifnum #1\expandafter \@firstoftwo
 \else \expandafter \@secondoftwo
 \fi
}%
\providecommand \@ifx [1]{%
 \ifx #1\expandafter \@firstoftwo
 \else \expandafter \@secondoftwo
 \fi
}%
\providecommand \natexlab [1]{#1}%
\providecommand \enquote  [1]{``#1''}%
\providecommand \bibnamefont  [1]{#1}%
\providecommand \bibfnamefont [1]{#1}%
\providecommand \citenamefont [1]{#1}%
\providecommand \href@noop [0]{\@secondoftwo}%
\providecommand \href [0]{\begingroup \@sanitize@url \@href}%
\providecommand \@href[1]{\@@startlink{#1}\@@href}%
\providecommand \@@href[1]{\endgroup#1\@@endlink}%
\providecommand \@sanitize@url [0]{\catcode `\\12\catcode `\$12\catcode
  `\&12\catcode `\#12\catcode `\^12\catcode `\_12\catcode `\%12\relax}%
\providecommand \@@startlink[1]{}%
\providecommand \@@endlink[0]{}%
\providecommand \url  [0]{\begingroup\@sanitize@url \@url }%
\providecommand \@url [1]{\endgroup\@href {#1}{\urlprefix }}%
\providecommand \urlprefix  [0]{URL }%
\providecommand \Eprint [0]{\href }%
\providecommand \doibase [0]{http://dx.doi.org/}%
\providecommand \selectlanguage [0]{\@gobble}%
\providecommand \bibinfo  [0]{\@secondoftwo}%
\providecommand \bibfield  [0]{\@secondoftwo}%
\providecommand \translation [1]{[#1]}%
\providecommand \BibitemOpen [0]{}%
\providecommand \bibitemStop [0]{}%
\providecommand \bibitemNoStop [0]{.\EOS\space}%
\providecommand \EOS [0]{\spacefactor3000\relax}%
\providecommand \BibitemShut  [1]{\csname bibitem#1\endcsname}%
\let\auto@bib@innerbib\@empty
%</preamble>
\bibitem [{\citenamefont {Jonsson}\ \emph {et~al.}(2015)\citenamefont
  {Jonsson}, \citenamefont {Martín-Martínez},\ and\ \citenamefont
  {Kempf}}]{jonsson_information_2015}%
  \BibitemOpen
  \bibfield  {author} {\bibinfo {author} {\bibfnamefont {R.~H.}\ \bibnamefont
  {Jonsson}}, \bibinfo {author} {\bibfnamefont {E.}~\bibnamefont
  {Martín-Martínez}}, \ and\ \bibinfo {author} {\bibfnamefont
  {A.}~\bibnamefont {Kempf}},\ }\href {\doibase 10.1103/PhysRevLett.114.110505}
  {\bibfield  {journal} {\bibinfo  {journal} {Physical Review Letters}\
  }\textbf {\bibinfo {volume} {114}},\ \bibinfo {pages} {110505} (\bibinfo
  {year} {2015})}\BibitemShut {NoStop}%
\bibitem [{\citenamefont {Ahmadzadegan}\ \emph {et~al.}(2018)\citenamefont
  {Ahmadzadegan}, \citenamefont {Martin-Martinez},\ and\ \citenamefont
  {Kempf}}]{ahmadzadegan_quantum_2018}%
  \BibitemOpen
  \bibfield  {author} {\bibinfo {author} {\bibfnamefont {A.}~\bibnamefont
  {Ahmadzadegan}}, \bibinfo {author} {\bibfnamefont {E.}~\bibnamefont
  {Martin-Martinez}}, \ and\ \bibinfo {author} {\bibfnamefont {A.}~\bibnamefont
  {Kempf}},\ }\href {http://arxiv.org/abs/1811.10606} {\bibfield  {journal}
  {\bibinfo  {journal} {arXiv:1811.10606 [quant-ph]}\ } (\bibinfo {year}
  {2018})}\BibitemShut {NoStop}%
\bibitem [{\citenamefont {Yamaguchi}\ \emph {et~al.}(2020)\citenamefont
  {Yamaguchi}, \citenamefont {Ahmadzadegan}, \citenamefont {Simidzija},
  \citenamefont {Kempf},\ and\ \citenamefont
  {Martín-Martínez}}]{yamaguchi_superadditivity_2020}%
  \BibitemOpen
  \bibfield  {author} {\bibinfo {author} {\bibfnamefont {K.}~\bibnamefont
  {Yamaguchi}}, \bibinfo {author} {\bibfnamefont {A.}~\bibnamefont
  {Ahmadzadegan}}, \bibinfo {author} {\bibfnamefont {P.}~\bibnamefont
  {Simidzija}}, \bibinfo {author} {\bibfnamefont {A.}~\bibnamefont {Kempf}}, \
  and\ \bibinfo {author} {\bibfnamefont {E.}~\bibnamefont
  {Martín-Martínez}},\ }\href {\doibase 10.1103/PhysRevD.101.105009}
  {\bibfield  {journal} {\bibinfo  {journal} {Physical Review D}\ }\textbf
  {\bibinfo {volume} {101}},\ \bibinfo {pages} {105009} (\bibinfo {year}
  {2020})}\BibitemShut {NoStop}%
\bibitem [{\citenamefont {Summers}\ and\ \citenamefont
  {Werner}(1985)}]{summers_vacuum_1985}%
  \BibitemOpen
  \bibfield  {author} {\bibinfo {author} {\bibfnamefont {S.~J.}\ \bibnamefont
  {Summers}}\ and\ \bibinfo {author} {\bibfnamefont {R.}~\bibnamefont
  {Werner}},\ }\href {\doibase 10.1016/0375-9601(85)90093-3} {\bibfield
  {journal} {\bibinfo  {journal} {Physics Letters A}\ }\textbf {\bibinfo
  {volume} {110}},\ \bibinfo {pages} {257} (\bibinfo {year}
  {1985})}\BibitemShut {NoStop}%
\bibitem [{\citenamefont {Summers}\ and\ \citenamefont
  {Werner}(1987)}]{summers_bells_1987}%
  \BibitemOpen
  \bibfield  {author} {\bibinfo {author} {\bibfnamefont {S.~J.}\ \bibnamefont
  {Summers}}\ and\ \bibinfo {author} {\bibfnamefont {R.}~\bibnamefont
  {Werner}},\ }\href {\doibase 10.1063/1.527733} {\bibfield  {journal}
  {\bibinfo  {journal} {Journal of Mathematical Physics}\ }\textbf {\bibinfo
  {volume} {28}},\ \bibinfo {pages} {2440} (\bibinfo {year}
  {1987})}\BibitemShut {NoStop}%
\bibitem [{\citenamefont {Ahmadzadegan}\ \emph {et~al.}(2021)\citenamefont
  {Ahmadzadegan}, \citenamefont {Simidzija}, \citenamefont {Li},\ and\
  \citenamefont {Kempf}}]{ahmadzadegan_neural_2021}%
  \BibitemOpen
  \bibfield  {author} {\bibinfo {author} {\bibfnamefont {A.}~\bibnamefont
  {Ahmadzadegan}}, \bibinfo {author} {\bibfnamefont {P.}~\bibnamefont
  {Simidzija}}, \bibinfo {author} {\bibfnamefont {M.}~\bibnamefont {Li}}, \
  and\ \bibinfo {author} {\bibfnamefont {A.}~\bibnamefont {Kempf}},\ }\href
  {\doibase 10.1038/s41598-021-00502-4} {\bibfield  {journal} {\bibinfo
  {journal} {Scientific Reports}\ }\textbf {\bibinfo {volume} {11}},\ \bibinfo
  {pages} {21624} (\bibinfo {year} {2021})}\BibitemShut {NoStop}%
\bibitem [{\citenamefont {Unruh}(1976)}]{unruh_notes_1976}%
  \BibitemOpen
  \bibfield  {author} {\bibinfo {author} {\bibfnamefont {W.~G.}\ \bibnamefont
  {Unruh}},\ }\href {\doibase 10.1103/PhysRevD.14.870} {\bibfield  {journal}
  {\bibinfo  {journal} {Physical Review D}\ }\textbf {\bibinfo {volume} {14}},\
  \bibinfo {pages} {870} (\bibinfo {year} {1976})}\BibitemShut {NoStop}%
\bibitem [{\citenamefont {DeWitt}(1979)}]{dewitt_quantum_1979}%
  \BibitemOpen
  \bibfield  {author} {\bibinfo {author} {\bibfnamefont {B.}~\bibnamefont
  {DeWitt}},\ }in\ \href@noop {} {\emph {\bibinfo {booktitle} {General
  relativity: an {Einstein} centenary survey}}}\ (\bibinfo  {publisher}
  {Cambridge University Press},\ \bibinfo {address} {New York},\ \bibinfo
  {year} {1979})\BibitemShut {NoStop}%
\bibitem [{\citenamefont {Valentini}(1991)}]{valentini_non-local_1991}%
  \BibitemOpen
  \bibfield  {author} {\bibinfo {author} {\bibfnamefont {A.}~\bibnamefont
  {Valentini}},\ }\href {\doibase 10.1016/0375-9601(91)90952-5} {\bibfield
  {journal} {\bibinfo  {journal} {Physics Letters A}\ }\textbf {\bibinfo
  {volume} {153}},\ \bibinfo {pages} {321} (\bibinfo {year}
  {1991})}\BibitemShut {NoStop}%
\bibitem [{\citenamefont {Reznik}(2003)}]{reznik_entanglement_2003}%
  \BibitemOpen
  \bibfield  {author} {\bibinfo {author} {\bibfnamefont {B.}~\bibnamefont
  {Reznik}},\ }\href {\doibase 10.1023/A:1022875910744} {\bibfield  {journal}
  {\bibinfo  {journal} {Foundations of Physics}\ }\textbf {\bibinfo {volume}
  {33}},\ \bibinfo {pages} {167} (\bibinfo {year} {2003})}\BibitemShut
  {NoStop}%
\bibitem [{\citenamefont {Reznik}\ \emph {et~al.}(2005)\citenamefont {Reznik},
  \citenamefont {Retzker},\ and\ \citenamefont
  {Silman}}]{reznik_violating_2005}%
  \BibitemOpen
  \bibfield  {author} {\bibinfo {author} {\bibfnamefont {B.}~\bibnamefont
  {Reznik}}, \bibinfo {author} {\bibfnamefont {A.}~\bibnamefont {Retzker}}, \
  and\ \bibinfo {author} {\bibfnamefont {J.}~\bibnamefont {Silman}},\ }\href
  {\doibase 10.1103/PhysRevA.71.042104} {\bibfield  {journal} {\bibinfo
  {journal} {Physical Review A}\ }\textbf {\bibinfo {volume} {71}},\ \bibinfo
  {pages} {042104} (\bibinfo {year} {2005})}\BibitemShut {NoStop}%
\bibitem [{\citenamefont {Retzker}\ \emph {et~al.}(2005)\citenamefont
  {Retzker}, \citenamefont {Cirac},\ and\ \citenamefont
  {Reznik}}]{retzker_detecting_2005}%
  \BibitemOpen
  \bibfield  {author} {\bibinfo {author} {\bibfnamefont {A.}~\bibnamefont
  {Retzker}}, \bibinfo {author} {\bibfnamefont {J.~I.}\ \bibnamefont {Cirac}},
  \ and\ \bibinfo {author} {\bibfnamefont {B.}~\bibnamefont {Reznik}},\ }\href
  {\doibase 10.1103/PhysRevLett.94.050504} {\bibfield  {journal} {\bibinfo
  {journal} {Physical Review Letters}\ }\textbf {\bibinfo {volume} {94}},\
  \bibinfo {pages} {050504} (\bibinfo {year} {2005})}\BibitemShut {NoStop}%
\bibitem [{\citenamefont {Silman}\ and\ \citenamefont
  {Reznik}(2007)}]{silman_long-range_2007}%
  \BibitemOpen
  \bibfield  {author} {\bibinfo {author} {\bibfnamefont {J.}~\bibnamefont
  {Silman}}\ and\ \bibinfo {author} {\bibfnamefont {B.}~\bibnamefont
  {Reznik}},\ }\href {\doibase 10.1103/PhysRevA.75.052307} {\bibfield
  {journal} {\bibinfo  {journal} {Physical Review A}\ }\textbf {\bibinfo
  {volume} {75}},\ \bibinfo {pages} {052307} (\bibinfo {year}
  {2007})}\BibitemShut {NoStop}%
\bibitem [{\citenamefont {Steeg}\ and\ \citenamefont
  {Menicucci}(2009)}]{steeg_entangling_2009}%
  \BibitemOpen
  \bibfield  {author} {\bibinfo {author} {\bibfnamefont {G.~V.}\ \bibnamefont
  {Steeg}}\ and\ \bibinfo {author} {\bibfnamefont {N.~C.}\ \bibnamefont
  {Menicucci}},\ }\href {\doibase 10.1103/PhysRevD.79.044027} {\bibfield
  {journal} {\bibinfo  {journal} {Physical Review D}\ }\textbf {\bibinfo
  {volume} {79}},\ \bibinfo {pages} {044027} (\bibinfo {year}
  {2009})}\BibitemShut {NoStop}%
\bibitem [{\citenamefont {Olson}\ and\ \citenamefont
  {Ralph}(2011)}]{olson_entanglement_2011}%
  \BibitemOpen
  \bibfield  {author} {\bibinfo {author} {\bibfnamefont {S.~J.}\ \bibnamefont
  {Olson}}\ and\ \bibinfo {author} {\bibfnamefont {T.~C.}\ \bibnamefont
  {Ralph}},\ }\href {\doibase 10.1103/PhysRevLett.106.110404} {\bibfield
  {journal} {\bibinfo  {journal} {Physical Review Letters}\ }\textbf {\bibinfo
  {volume} {106}},\ \bibinfo {pages} {110404} (\bibinfo {year}
  {2011})}\BibitemShut {NoStop}%
\bibitem [{\citenamefont {Olson}\ and\ \citenamefont
  {Ralph}(2012)}]{olson_extraction_2012}%
  \BibitemOpen
  \bibfield  {author} {\bibinfo {author} {\bibfnamefont {S.~J.}\ \bibnamefont
  {Olson}}\ and\ \bibinfo {author} {\bibfnamefont {T.~C.}\ \bibnamefont
  {Ralph}},\ }\href {\doibase 10.1103/PhysRevA.85.012306} {\bibfield  {journal}
  {\bibinfo  {journal} {Physical Review A}\ }\textbf {\bibinfo {volume} {85}},\
  \bibinfo {pages} {012306} (\bibinfo {year} {2012})}\BibitemShut {NoStop}%
\bibitem [{\citenamefont {Pozas-Kerstjens}\ and\ \citenamefont
  {Martín-Martínez}(2015)}]{pozas-kerstjens_harvesting_2015}%
  \BibitemOpen
  \bibfield  {author} {\bibinfo {author} {\bibfnamefont {A.}~\bibnamefont
  {Pozas-Kerstjens}}\ and\ \bibinfo {author} {\bibfnamefont {E.}~\bibnamefont
  {Martín-Martínez}},\ }\href {\doibase 10.1103/PhysRevD.92.064042}
  {\bibfield  {journal} {\bibinfo  {journal} {Physical Review D}\ }\textbf
  {\bibinfo {volume} {92}},\ \bibinfo {pages} {064042} (\bibinfo {year}
  {2015})}\BibitemShut {NoStop}%
\bibitem [{\citenamefont {Pozas-Kerstjens}\ and\ \citenamefont
  {Martín-Martínez}(2016)}]{pozas-kerstjens_entanglement_2016}%
  \BibitemOpen
  \bibfield  {author} {\bibinfo {author} {\bibfnamefont {A.}~\bibnamefont
  {Pozas-Kerstjens}}\ and\ \bibinfo {author} {\bibfnamefont {E.}~\bibnamefont
  {Martín-Martínez}},\ }\href {\doibase 10.1103/PhysRevD.94.064074}
  {\bibfield  {journal} {\bibinfo  {journal} {Physical Review D}\ }\textbf
  {\bibinfo {volume} {94}},\ \bibinfo {pages} {064074} (\bibinfo {year}
  {2016})}\BibitemShut {NoStop}%
\bibitem [{\citenamefont {Salton}\ \emph {et~al.}(2015)\citenamefont {Salton},
  \citenamefont {Mann},\ and\ \citenamefont
  {Menicucci}}]{salton_acceleration-assisted_2015}%
  \BibitemOpen
  \bibfield  {author} {\bibinfo {author} {\bibfnamefont {G.}~\bibnamefont
  {Salton}}, \bibinfo {author} {\bibfnamefont {R.~B.}\ \bibnamefont {Mann}}, \
  and\ \bibinfo {author} {\bibfnamefont {N.~C.}\ \bibnamefont {Menicucci}},\
  }\href {\doibase 10.1088/1367-2630/17/3/035001} {\bibfield  {journal}
  {\bibinfo  {journal} {New Journal of Physics}\ }\textbf {\bibinfo {volume}
  {17}},\ \bibinfo {pages} {035001} (\bibinfo {year} {2015})}\BibitemShut
  {NoStop}%
\bibitem [{\citenamefont {Nambu}(2013)}]{nambu_entanglement_2013}%
  \BibitemOpen
  \bibfield  {author} {\bibinfo {author} {\bibfnamefont {Y.}~\bibnamefont
  {Nambu}},\ }\href {\doibase 10.3390/e15051847} {\bibfield  {journal}
  {\bibinfo  {journal} {Entropy}\ }\textbf {\bibinfo {volume} {15}},\ \bibinfo
  {pages} {1847} (\bibinfo {year} {2013})}\BibitemShut {NoStop}%
\bibitem [{\citenamefont {Sabín}\ \emph {et~al.}(2012)\citenamefont {Sabín},
  \citenamefont {Peropadre}, \citenamefont {del Rey},\ and\ \citenamefont
  {Martín-Martínez}}]{sabin_extracting_2012}%
  \BibitemOpen
  \bibfield  {author} {\bibinfo {author} {\bibfnamefont {C.}~\bibnamefont
  {Sabín}}, \bibinfo {author} {\bibfnamefont {B.}~\bibnamefont {Peropadre}},
  \bibinfo {author} {\bibfnamefont {M.}~\bibnamefont {del Rey}}, \ and\
  \bibinfo {author} {\bibfnamefont {E.}~\bibnamefont {Martín-Martínez}},\
  }\href {\doibase 10.1103/PhysRevLett.109.033602} {\bibfield  {journal}
  {\bibinfo  {journal} {Physical Review Letters}\ }\textbf {\bibinfo {volume}
  {109}},\ \bibinfo {pages} {033602} (\bibinfo {year} {2012})}\BibitemShut
  {NoStop}%
\bibitem [{\citenamefont {Martín-Martínez}\ \emph {et~al.}(2013)\citenamefont
  {Martín-Martínez}, \citenamefont {Brown}, \citenamefont {Donnelly},\ and\
  \citenamefont {Kempf}}]{martin-martinez_sustainable_2013}%
  \BibitemOpen
  \bibfield  {author} {\bibinfo {author} {\bibfnamefont {E.}~\bibnamefont
  {Martín-Martínez}}, \bibinfo {author} {\bibfnamefont {E.~G.}\ \bibnamefont
  {Brown}}, \bibinfo {author} {\bibfnamefont {W.}~\bibnamefont {Donnelly}}, \
  and\ \bibinfo {author} {\bibfnamefont {A.}~\bibnamefont {Kempf}},\ }\href
  {\doibase 10.1103/PhysRevA.88.052310} {\bibfield  {journal} {\bibinfo
  {journal} {Physical Review A}\ }\textbf {\bibinfo {volume} {88}},\ \bibinfo
  {pages} {052310} (\bibinfo {year} {2013})}\BibitemShut {NoStop}%
\bibitem [{\citenamefont {Lin}\ and\ \citenamefont
  {Hu}(2010)}]{lin_entanglement_2010}%
  \BibitemOpen
  \bibfield  {author} {\bibinfo {author} {\bibfnamefont {S.-Y.}\ \bibnamefont
  {Lin}}\ and\ \bibinfo {author} {\bibfnamefont {B.~L.}\ \bibnamefont {Hu}},\
  }\href {\doibase 10.1103/PhysRevD.81.045019} {\bibfield  {journal} {\bibinfo
  {journal} {Physical Review D}\ }\textbf {\bibinfo {volume} {81}},\ \bibinfo
  {pages} {045019} (\bibinfo {year} {2010})}\BibitemShut {NoStop}%
\bibitem [{\citenamefont {Kukita}\ and\ \citenamefont
  {Nambu}(2017{\natexlab{a}})}]{kukita_entanglement_2017}%
  \BibitemOpen
  \bibfield  {author} {\bibinfo {author} {\bibfnamefont {S.}~\bibnamefont
  {Kukita}}\ and\ \bibinfo {author} {\bibfnamefont {Y.}~\bibnamefont {Nambu}},\
  }\href {\doibase 10.1088/1361-6382/aa8e31} {\bibfield  {journal} {\bibinfo
  {journal} {Classical and Quantum Gravity}\ }\textbf {\bibinfo {volume}
  {34}},\ \bibinfo {pages} {235010} (\bibinfo {year}
  {2017}{\natexlab{a}})}\BibitemShut {NoStop}%
\bibitem [{\citenamefont {Kukita}\ and\ \citenamefont
  {Nambu}(2017{\natexlab{b}})}]{kukita_harvesting_2017}%
  \BibitemOpen
  \bibfield  {author} {\bibinfo {author} {\bibfnamefont {S.}~\bibnamefont
  {Kukita}}\ and\ \bibinfo {author} {\bibfnamefont {Y.}~\bibnamefont {Nambu}},\
  }\href {\doibase 10.3390/e19090449} {\bibfield  {journal} {\bibinfo
  {journal} {Entropy}\ }\textbf {\bibinfo {volume} {19}},\ \bibinfo {pages}
  {449} (\bibinfo {year} {2017}{\natexlab{b}})}\BibitemShut {NoStop}%
\bibitem [{\citenamefont {Ng}\ \emph {et~al.}(2018)\citenamefont {Ng},
  \citenamefont {Mann},\ and\ \citenamefont
  {Martín-Martínez}}]{ng_unruh-dewitt_2018}%
  \BibitemOpen
  \bibfield  {author} {\bibinfo {author} {\bibfnamefont {K.~K.}\ \bibnamefont
  {Ng}}, \bibinfo {author} {\bibfnamefont {R.~B.}\ \bibnamefont {Mann}}, \ and\
  \bibinfo {author} {\bibfnamefont {E.}~\bibnamefont {Martín-Martínez}},\
  }\href {\doibase 10.1103/PhysRevD.98.125005} {\bibfield  {journal} {\bibinfo
  {journal} {Physical Review D}\ }\textbf {\bibinfo {volume} {98}},\ \bibinfo
  {pages} {125005} (\bibinfo {year} {2018})}\BibitemShut {NoStop}%
\bibitem [{\citenamefont {Henderson}\ \emph {et~al.}(2018)\citenamefont
  {Henderson}, \citenamefont {Hennigar}, \citenamefont {Mann}, \citenamefont
  {Smith},\ and\ \citenamefont {Zhang}}]{henderson_harvesting_2018}%
  \BibitemOpen
  \bibfield  {author} {\bibinfo {author} {\bibfnamefont {L.~J.}\ \bibnamefont
  {Henderson}}, \bibinfo {author} {\bibfnamefont {R.~A.}\ \bibnamefont
  {Hennigar}}, \bibinfo {author} {\bibfnamefont {R.~B.}\ \bibnamefont {Mann}},
  \bibinfo {author} {\bibfnamefont {A.~R.~H.}\ \bibnamefont {Smith}}, \ and\
  \bibinfo {author} {\bibfnamefont {J.}~\bibnamefont {Zhang}},\ }\href
  {\doibase 10.1088/1361-6382/aae27e} {\bibfield  {journal} {\bibinfo
  {journal} {Classical and Quantum Gravity}\ }\textbf {\bibinfo {volume}
  {35}},\ \bibinfo {pages} {21LT02} (\bibinfo {year} {2018})}\BibitemShut
  {NoStop}%
\bibitem [{\citenamefont {Henderson}\ \emph {et~al.}(2019)\citenamefont
  {Henderson}, \citenamefont {Hennigar}, \citenamefont {Mann}, \citenamefont
  {Smith},\ and\ \citenamefont {Zhang}}]{henderson_entangling_2019}%
  \BibitemOpen
  \bibfield  {author} {\bibinfo {author} {\bibfnamefont {L.~J.}\ \bibnamefont
  {Henderson}}, \bibinfo {author} {\bibfnamefont {R.~A.}\ \bibnamefont
  {Hennigar}}, \bibinfo {author} {\bibfnamefont {R.~B.}\ \bibnamefont {Mann}},
  \bibinfo {author} {\bibfnamefont {A.~R.~H.}\ \bibnamefont {Smith}}, \ and\
  \bibinfo {author} {\bibfnamefont {J.}~\bibnamefont {Zhang}},\ }\href
  {\doibase 10.1007/JHEP05(2019)178} {\bibfield  {journal} {\bibinfo  {journal}
  {Journal of High Energy Physics}\ }\textbf {\bibinfo {volume} {2019}},\
  \bibinfo {pages} {178} (\bibinfo {year} {2019})}\BibitemShut {NoStop}%
\bibitem [{\citenamefont {Simidzija}\ and\ \citenamefont
  {Martín-Martínez}(2018)}]{simidzija_harvesting_2018}%
  \BibitemOpen
  \bibfield  {author} {\bibinfo {author} {\bibfnamefont {P.}~\bibnamefont
  {Simidzija}}\ and\ \bibinfo {author} {\bibfnamefont {E.}~\bibnamefont
  {Martín-Martínez}},\ }\href {\doibase 10.1103/PhysRevD.98.085007}
  {\bibfield  {journal} {\bibinfo  {journal} {Physical Review D}\ }\textbf
  {\bibinfo {volume} {98}},\ \bibinfo {pages} {085007} (\bibinfo {year}
  {2018})}\BibitemShut {NoStop}%
\bibitem [{\citenamefont {Braun}(2002)}]{braun_creation_2002}%
  \BibitemOpen
  \bibfield  {author} {\bibinfo {author} {\bibfnamefont {D.}~\bibnamefont
  {Braun}},\ }\href {\doibase 10.1103/PhysRevLett.89.277901} {\bibfield
  {journal} {\bibinfo  {journal} {Physical Review Letters}\ }\textbf {\bibinfo
  {volume} {89}},\ \bibinfo {pages} {277901} (\bibinfo {year}
  {2002})}\BibitemShut {NoStop}%
\bibitem [{\citenamefont {Braun}(2005)}]{braun_entanglement_2005}%
  \BibitemOpen
  \bibfield  {author} {\bibinfo {author} {\bibfnamefont {D.}~\bibnamefont
  {Braun}},\ }\href {\doibase 10.1103/PhysRevA.72.062324} {\bibfield  {journal}
  {\bibinfo  {journal} {Physical Review A}\ }\textbf {\bibinfo {volume} {72}},\
  \bibinfo {pages} {062324} (\bibinfo {year} {2005})}\BibitemShut {NoStop}%
\bibitem [{\citenamefont {Hotta}\ \emph {et~al.}(2020)\citenamefont {Hotta},
  \citenamefont {Kempf}, \citenamefont {Martín-Martínez}, \citenamefont
  {Tomitsuka},\ and\ \citenamefont {Yamaguchi}}]{hotta_duality_2020}%
  \BibitemOpen
  \bibfield  {author} {\bibinfo {author} {\bibfnamefont {M.}~\bibnamefont
  {Hotta}}, \bibinfo {author} {\bibfnamefont {A.}~\bibnamefont {Kempf}},
  \bibinfo {author} {\bibfnamefont {E.}~\bibnamefont {Martín-Martínez}},
  \bibinfo {author} {\bibfnamefont {T.}~\bibnamefont {Tomitsuka}}, \ and\
  \bibinfo {author} {\bibfnamefont {K.}~\bibnamefont {Yamaguchi}},\ }\href
  {\doibase 10.1103/PhysRevD.101.085017} {\bibfield  {journal} {\bibinfo
  {journal} {Physical Review D}\ }\textbf {\bibinfo {volume} {101}},\ \bibinfo
  {pages} {085017} (\bibinfo {year} {2020})}\BibitemShut {NoStop}%
\bibitem [{\citenamefont {Martín-Martínez}\ \emph {et~al.}(2016)\citenamefont
  {Martín-Martínez}, \citenamefont {Smith},\ and\ \citenamefont
  {Terno}}]{martin-martinez_spacetime_2016}%
  \BibitemOpen
  \bibfield  {author} {\bibinfo {author} {\bibfnamefont {E.}~\bibnamefont
  {Martín-Martínez}}, \bibinfo {author} {\bibfnamefont {A.~R.}\ \bibnamefont
  {Smith}}, \ and\ \bibinfo {author} {\bibfnamefont {D.~R.}\ \bibnamefont
  {Terno}},\ }\href {\doibase 10.1103/PhysRevD.93.044001} {\bibfield  {journal}
  {\bibinfo  {journal} {Physical Review D}\ }\textbf {\bibinfo {volume} {93}},\
  \bibinfo {pages} {044001} (\bibinfo {year} {2016})}\BibitemShut {NoStop}%
\bibitem [{\citenamefont {Saravani}\ \emph {et~al.}(2016)\citenamefont
  {Saravani}, \citenamefont {Aslanbeigi},\ and\ \citenamefont
  {Kempf}}]{saravani_spacetime_2016}%
  \BibitemOpen
  \bibfield  {author} {\bibinfo {author} {\bibfnamefont {M.}~\bibnamefont
  {Saravani}}, \bibinfo {author} {\bibfnamefont {S.}~\bibnamefont
  {Aslanbeigi}}, \ and\ \bibinfo {author} {\bibfnamefont {A.}~\bibnamefont
  {Kempf}},\ }\href {\doibase 10.1103/PhysRevD.93.045026} {\bibfield  {journal}
  {\bibinfo  {journal} {Physical Review D}\ }\textbf {\bibinfo {volume} {93}},\
  \bibinfo {pages} {045026} (\bibinfo {year} {2016})}\BibitemShut {NoStop}%
\bibitem [{\citenamefont {Kempf}(2021)}]{kempf_replacing_2021}%
  \BibitemOpen
  \bibfield  {author} {\bibinfo {author} {\bibfnamefont {A.}~\bibnamefont
  {Kempf}},\ }\href
  {https://www.frontiersin.org/articles/10.3389/fphy.2021.655857} {\bibfield
  {journal} {\bibinfo  {journal} {Frontiers in Physics}\ }\textbf {\bibinfo
  {volume} {9}} (\bibinfo {year} {2021})}\BibitemShut {NoStop}%
\bibitem [{\citenamefont {Perche}\ and\ \citenamefont
  {Martín-Martínez}(2022)}]{perche_geometry_2022}%
  \BibitemOpen
  \bibfield  {author} {\bibinfo {author} {\bibfnamefont {T.~R.}\ \bibnamefont
  {Perche}}\ and\ \bibinfo {author} {\bibfnamefont {E.}~\bibnamefont
  {Martín-Martínez}},\ }\href {\doibase 10.1103/PhysRevD.105.066011}
  {\bibfield  {journal} {\bibinfo  {journal} {Physical Review D}\ }\textbf
  {\bibinfo {volume} {105}},\ \bibinfo {pages} {066011} (\bibinfo {year}
  {2022})}\BibitemShut {NoStop}%
\bibitem [{\citenamefont {Cliche}\ and\ \citenamefont
  {Kempf}(2010)}]{cliche_relativistic_2010}%
  \BibitemOpen
  \bibfield  {author} {\bibinfo {author} {\bibfnamefont {M.}~\bibnamefont
  {Cliche}}\ and\ \bibinfo {author} {\bibfnamefont {A.}~\bibnamefont {Kempf}},\
  }\href {\doibase 10.1103/PhysRevA.81.012330} {\bibfield  {journal} {\bibinfo
  {journal} {Physical Review A}\ }\textbf {\bibinfo {volume} {81}},\ \bibinfo
  {pages} {012330} (\bibinfo {year} {2010})}\BibitemShut {NoStop}%
\bibitem [{\citenamefont {Koga}\ \emph {et~al.}(2018)\citenamefont {Koga},
  \citenamefont {Kimura},\ and\ \citenamefont {Maeda}}]{koga_quantum_2018}%
  \BibitemOpen
  \bibfield  {author} {\bibinfo {author} {\bibfnamefont {J.-i.}\ \bibnamefont
  {Koga}}, \bibinfo {author} {\bibfnamefont {G.}~\bibnamefont {Kimura}}, \ and\
  \bibinfo {author} {\bibfnamefont {K.}~\bibnamefont {Maeda}},\ }\href
  {\doibase 10.1103/PhysRevA.97.062338} {\bibfield  {journal} {\bibinfo
  {journal} {Physical Review A}\ }\textbf {\bibinfo {volume} {97}},\ \bibinfo
  {pages} {062338} (\bibinfo {year} {2018})}\BibitemShut {NoStop}%
\bibitem [{\citenamefont {Apollaro}\ \emph {et~al.}(2023)\citenamefont
  {Apollaro}, \citenamefont {Lorenzo}, \citenamefont {Plastina}, \citenamefont
  {Consiglio},\ and\ \citenamefont {Życzkowski}}]{apollaro_entangled_2023}%
  \BibitemOpen
  \bibfield  {author} {\bibinfo {author} {\bibfnamefont {T.~J.~G.}\
  \bibnamefont {Apollaro}}, \bibinfo {author} {\bibfnamefont {S.}~\bibnamefont
  {Lorenzo}}, \bibinfo {author} {\bibfnamefont {F.}~\bibnamefont {Plastina}},
  \bibinfo {author} {\bibfnamefont {M.}~\bibnamefont {Consiglio}}, \ and\
  \bibinfo {author} {\bibfnamefont {K.}~\bibnamefont {Życzkowski}},\ }\href
  {\doibase 10.3390/e25010046} {\bibfield  {journal} {\bibinfo  {journal}
  {Entropy}\ }\textbf {\bibinfo {volume} {25}},\ \bibinfo {pages} {46}
  (\bibinfo {year} {2023})}\BibitemShut {NoStop}%
\bibitem [{\citenamefont {Bennett}\ \emph {et~al.}(1993)\citenamefont
  {Bennett}, \citenamefont {Brassard}, \citenamefont {Crépeau}, \citenamefont
  {Jozsa}, \citenamefont {Peres},\ and\ \citenamefont
  {Wootters}}]{bennett_teleporting_1993}%
  \BibitemOpen
  \bibfield  {author} {\bibinfo {author} {\bibfnamefont {C.~H.}\ \bibnamefont
  {Bennett}}, \bibinfo {author} {\bibfnamefont {G.}~\bibnamefont {Brassard}},
  \bibinfo {author} {\bibfnamefont {C.}~\bibnamefont {Crépeau}}, \bibinfo
  {author} {\bibfnamefont {R.}~\bibnamefont {Jozsa}}, \bibinfo {author}
  {\bibfnamefont {A.}~\bibnamefont {Peres}}, \ and\ \bibinfo {author}
  {\bibfnamefont {W.~K.}\ \bibnamefont {Wootters}},\ }\href {\doibase
  10.1103/PhysRevLett.70.1895} {\bibfield  {journal} {\bibinfo  {journal}
  {Physical Review Letters}\ }\textbf {\bibinfo {volume} {70}},\ \bibinfo
  {pages} {1895} (\bibinfo {year} {1993})}\BibitemShut {NoStop}%
\bibitem [{\citenamefont {Vidal}\ and\ \citenamefont
  {Werner}(2002)}]{vidal_computable_2002}%
  \BibitemOpen
  \bibfield  {author} {\bibinfo {author} {\bibfnamefont {G.}~\bibnamefont
  {Vidal}}\ and\ \bibinfo {author} {\bibfnamefont {R.~F.}\ \bibnamefont
  {Werner}},\ }\href {\doibase 10.1103/PhysRevA.65.032314} {\bibfield
  {journal} {\bibinfo  {journal} {Physical Review A}\ }\textbf {\bibinfo
  {volume} {65}},\ \bibinfo {pages} {032314} (\bibinfo {year}
  {2002})}\BibitemShut {NoStop}%
\bibitem [{\citenamefont {Horodecki}\ \emph {et~al.}(1996)\citenamefont
  {Horodecki}, \citenamefont {Horodecki},\ and\ \citenamefont
  {Horodecki}}]{horodecki_separability_1996}%
  \BibitemOpen
  \bibfield  {author} {\bibinfo {author} {\bibfnamefont {M.}~\bibnamefont
  {Horodecki}}, \bibinfo {author} {\bibfnamefont {P.}~\bibnamefont
  {Horodecki}}, \ and\ \bibinfo {author} {\bibfnamefont {R.}~\bibnamefont
  {Horodecki}},\ }\href {\doibase 10.1016/S0375-9601(96)00706-2} {\bibfield
  {journal} {\bibinfo  {journal} {Physics Letters A}\ }\textbf {\bibinfo
  {volume} {223}},\ \bibinfo {pages} {1} (\bibinfo {year} {1996})}\BibitemShut
  {NoStop}%
\bibitem [{\citenamefont {Wen}\ and\ \citenamefont
  {Kempf}(2022)}]{wen_transfer_2022}%
  \BibitemOpen
  \bibfield  {author} {\bibinfo {author} {\bibfnamefont {R.~Y.}\ \bibnamefont
  {Wen}}\ and\ \bibinfo {author} {\bibfnamefont {A.}~\bibnamefont {Kempf}},\
  }\href {\doibase 10.1088/1751-8121/aca7a1} {\bibfield  {journal} {\bibinfo
  {journal} {Journal of Physics A: Mathematical and Theoretical}\ }\textbf
  {\bibinfo {volume} {55}},\ \bibinfo {pages} {495304} (\bibinfo {year}
  {2022})}\BibitemShut {NoStop}%
\bibitem [{\citenamefont {Martín-Martínez}\ and\ \citenamefont
  {Rodriguez-Lopez}(2018)}]{martin-martinez_relativistic_2018}%
  \BibitemOpen
  \bibfield  {author} {\bibinfo {author} {\bibfnamefont {E.}~\bibnamefont
  {Martín-Martínez}}\ and\ \bibinfo {author} {\bibfnamefont {P.}~\bibnamefont
  {Rodriguez-Lopez}},\ }\href {\doibase 10.1103/PhysRevD.97.105026} {\bibfield
  {journal} {\bibinfo  {journal} {Physical Review D}\ }\textbf {\bibinfo
  {volume} {97}},\ \bibinfo {pages} {105026} (\bibinfo {year}
  {2018})}\BibitemShut {NoStop}%
\bibitem [{\citenamefont {Panine}\ and\ \citenamefont
  {Kempf}(2017)}]{panine_convexity_2017}%
  \BibitemOpen
  \bibfield  {author} {\bibinfo {author} {\bibfnamefont {M.}~\bibnamefont
  {Panine}}\ and\ \bibinfo {author} {\bibfnamefont {A.}~\bibnamefont {Kempf}},\
  }\href {\doibase 10.1142/S0219887817501572} {\bibfield  {journal} {\bibinfo
  {journal} {International Journal of Geometric Methods in Modern Physics}\
  }\textbf {\bibinfo {volume} {14}},\ \bibinfo {pages} {1750157} (\bibinfo
  {year} {2017})},\ \bibinfo {note} {arXiv:1607.00396 [math-ph]}\BibitemShut
  {NoStop}%
\bibitem [{\citenamefont {Kato}(1995)}]{kato_perturbation_1995}%
  \BibitemOpen
  \bibfield  {author} {\bibinfo {author} {\bibfnamefont {T.}~\bibnamefont
  {Kato}},\ }\href {http://link.springer.com/10.1007/978-3-642-66282-9} {\emph
  {\bibinfo {title} {Perturbation {Theory} for {Linear} {Operators}}}},\
  \bibinfo {series} {Classics in {Mathematics}}, Vol.\ \bibinfo {volume} {132}\
  (\bibinfo  {publisher} {Springer Berlin Heidelberg},\ \bibinfo {address}
  {Berlin, Heidelberg},\ \bibinfo {year} {1995})\BibitemShut {NoStop}%
\end{thebibliography}%

\newpage
\appendix
\section{Perturbation of the eigenvalues}\label{sec:appendix_perturbation_EV}

Here we briefly summarize known results on perturbation of eigenvalues that can be found e.g., in \cite{panine_convexity_2017,wen_transfer_2022}. For more comprehensive and rigorous arguments, see e.g., \cite{kato_perturbation_1995}.

In this section, we summarize a perturbative method for calculating the eigenvalues of a Hermitian operator. Let $S(t)$ be an Hermitian operator, which can be expanded as
\begin{align}
    S(t)=S^{(0)}+tS^{(1)}+t^2S^{(2)}+\mathcal{O}(t^3)
\end{align}
as $t\to 0$. In the main text, $S(t)$ corresponds to the partial transpose of the density operator $\rho_{\tilde{A}B}^{\top_{\widetilde{A}}}(\lambda)$, where $\lambda$ is the coupling constant. 

Let $\sigma(S^{(0)})$ denote the set of different eigenvalues of $S^{(0)}$. 
The operator is decomposed as
\begin{align}
    S^{(0)}=\sum_{s^{(0)}\in\sigma(S^{(0)})} s^{(0)}\Pi_{s^{(0)}},
\end{align}
where $\Pi_{s^{(0)}}$ denotes the projector on the eigenspace of $S^{(0)}$ with eigenvalue $s^{(0)}$. 

The eigenvectors and the eigenvalues of $S(t)$ are expanded as
\begin{align}
    \ket{\psi_{s_i}(t)}&=\ket{\psi_{s_i}^{(0)}}+t\ket{\psi_{s_i}^{(1)}}+t^2\ket{\psi_{s_i}^{(2)}}+\mathcal{O}(t^3)\\
    s_{i}(t)&=s^{(0)}+ts_{i}^{(1)}+t^2s_{i}^{(2)}+\mathcal{O}(t^3),
\end{align}
where $i$ is the label for the degeneracy of the eigenspace of $S^{(0)}$. We assume that $\{\ket{\psi_{s_i}^{(0)}}\}_i$ forms an orthornormal basis for the eigenspace of $S^{(0)}$ with eigenvalue $s^{(0)}$, i.e.,
\begin{align}
    \braket{\psi_{s_j}^{(0)}|\psi_{s_i}^{(0)}}=\delta_{ij},\quad \Pi_{s^{(0)}}=\sum_i\ket{\psi_{s_i}^{(0)}}\bra{\psi_{s_i}^{(0)}}.
\end{align}

From the eigenvalue equation, we have
\begin{align}
    &\left(S^{(0)}-s^{(0)}\right)\ket{\psi_{s_i}^{(0)}}=0\label{eq:0th}\\
    &\left(S^{(1)}-s_{i}^{(1)}\right)\ket{\psi_{s_i}^{(0)}}+\left(S^{(0)}-s^{(0)}\right)\ket{\psi_{s_i}^{(1)}}=0\label{eq:1st}\\
    &\left(S^{(2)}-s_{i}^{(2)}\right)\ket{\psi_{s_i}^{(0)}}+\left(S^{(1)}-s_{i}^{(1)}\right)\ket{\psi_{s_i}^{(1)}}\nonumber\\
    &\quad \quad \quad \quad \quad \quad \quad \quad \quad +\left(S^{(0)}-s^{(0)}\right)\ket{\psi_{s_i}^{(2)}}=0\label{eq:2nd}.
\end{align}

From Eq.~\eqref{eq:1st}, we have
\begin{align}
    \Braket{\psi_{s_i}^{(0)}|S^{(1)}|\psi_{s_j}^{(0)}}=s_i^{(1)}\delta_{ij},
\end{align}
which implies that $\{s_{i}^{(1)}\}_i$ are the eigenvalues of an Hermitian operator $\Pi_{s^{(0)}} S^{(1)}\Pi_{s^{(0)}}$ that is restricted on the eigenspace of $S^{(0)}$ associated with the eigenvalue $s^{(0)}$. In addition, $\{\ket{\psi_{s_i}^{(1)}}\}_i$ are the corresponding eigenvectors. 
Furthermore, we have
\begin{align}
  \Braket{\psi_{s_j'}^{(0)}|S^{(1)}-s_i^{(1)}|\psi_{s_i}^{(0)}}+\left(s^{(0)\prime}-s^{(0)}\right)\Braket{\psi_{s'_j}^{(0)}|\psi_{s_i}^{(1)}}=0
\end{align}
for $s^{(0)\prime}\in\sigma(S^{(0)})\setminus\{s^{(0)}\}$.
Therefore, we have
\begin{align}
    &\ket{\psi_{s_i}^{(1)}}=\sum_{i}c_{s_i}^{(1)}\ket{ \psi_{s_i}^{(0)}}\nonumber\\
    &-\sum_{s^{(0)\prime}\in\sigma(S^{(0)})\setminus\{s^{(0)}\}}\sum_{j}\frac{\Braket{\psi_{s'_j}^{(0)}|S^{(1)}|\psi_{s_i}^{(0)}}}{s^{(0)\prime}-s^{(0)}}\ket{\psi_{s'_j}^{(0)}},
\end{align}
where $c_{s_i}^{(1)}\in\mathbb{C}$ are some coefficients, which are not important for our purpose. 

From Eq.~\eqref{eq:2nd}, we get
\begin{align}
    &\Braket{\psi_{s_j}^{(0)}|S^{(2)}|\psi_{s_i}^{(0)}}-s_{i}^{(2)}\delta_{ij}\nonumber\\
    &\quad +\Braket{\psi_{s_j}^{(0)}|\left(S^{(1)}-s_{i}^{(1)}\right)|\psi_{s_i}^{(1)}}=0.
\end{align}
Since $\{\ket{\psi_{s_i}^{(0)}}\}_i$ are eigenvectors of $\Pi_{s^{(0)}}S^{(1)}\Pi_{s^{(0)}}$, we get
\begin{widetext}
\begin{align}
    \Braket{\psi_{s_j}^{(0)}|\left(S^{(1)}-s_{i}^{(1)}\right)|\psi_{s_i}^{(1)}}=-\sum_{s^{(0)\prime}\in\sigma(S^{(0)})\setminus\{s^{(0)}\}}\sum_{j}\frac{\Braket{\psi_{s_j}^{(0)}|S^{(1)}|\psi_{s'_j}^{(0)}}\Braket{\psi_{s'_j}^{(0)}|S^{(1)}|\psi_{s_i}^{(0)}}}{s^{(0)\prime}-s^{(0)}}.
\end{align}
Therefore,
\begin{align}
    s_{i}^{(2)}\delta_{ij}=\Braket{\psi_{s_j}^{(0)}|S^{(2)}-S^{(1)}\sum_{s^{(0)\prime}\in\sigma(S^{(0)})\setminus\{s^{(0)}\}}\frac{\Pi_{s^{(0)\prime}}}{s^{(0)\prime}-s^{(0)}}S^{(1)}|\psi_{s_i}^{(0)}}.
\end{align}
\end{widetext}
This equation implies that $\{s_{i}^{(2)}\}_{i}$ are the eigenvalues of 
\begin{align}
    \Pi_{s^{(0)}}\left(S^{(2)}-S^{(1)}\sum_{s^{(0)\prime}\in\sigma(S^{(0)})\setminus\{s^{(0)}\}}\frac{\Pi_{s^{(0)\prime}}}{s^{(0)\prime}-s^{(0)}}S^{(1)}\right)\Pi_{s^{(0)}}
\end{align}
that is restricted on the eigenspace of $S^{(0)}$ with eigenvalue $s^{(0)}$.

\section{A formula for the partial transpose}\label{sec:appendix_commute}
To prove Eq.~\eqref{eq:partial_transpose_formula}, let us first consider the case where $O_{AB}=O_A\otimes O_B$. In this case, we have
\begin{align}
     &\left(O_{AB}\mathbb{I}_A\otimes Y_B\right)^{\top_A}\nonumber\\
     &=\left(O_A\otimes  O_{B} Y_B\right)^{\top_A}\nonumber\\
     &=O_A^\top\otimes  O_{B} Y_B
     \nonumber\\
     &=\left(O_{AB}^{\top_{A}}\right)\mathbb{I}_A\otimes Y_B.
\end{align}
Similarly, $\left(\mathbb{I}_A\otimes Y_BO_{AB}\right)^{\top_A}=\mathbb{I}_A\otimes Y_B\left(O_{AB}^{\top_A}\right)$ holds.
Now, note that any linear operator is expanded as $O_{AB}=\sum_i O_A^{(i)}\otimes O_B^{(i)}$. With the linearity of the partial transpose operation, we complete the proof of Eq.~\eqref{eq:partial_transpose_formula}. 

\section{Derivation of Eq.~\eqref{eq:xi}}\label{sec:appendix_derivation_of_density_mat}
We here explain the detailed derivation of Eq.~\eqref{eq:xi}. 
The initial state of the total system is given by
\begin{align}
    \ket{\Psi}\bra{\Psi}_{\widetilde{A}A}\otimes \rho_{A'B},
\end{align}
where $\rho_{A'B}$ is the state of detectors after the entanglement harvesting, given in Eq.~\eqref{eq:second_order_eh_state}.

To implement quantum teleportation by consuming the entanglement resource $\rho_{A'B}$, Alice performs the Bell measurement on $AA'$. The unnormalized selective post-measurement state for $\widetilde{A}B$ is given by
\begin{align}
    \xi_{\widetilde{A}B}(\mu)=\mathrm{Tr}_{AA'}\left(M_\mu\ket{\Psi}\bra{\Psi}_{\widetilde{A}A}\otimes\rho_{A'B} M_\mu^\dag\right),
\end{align}
where 
\begin{align}
    M_\mu&\coloneqq \mathbb{I}_{\widetilde{A}}\otimes \ket{\Psi_\mu}\bra{\Psi_\mu}_{AA'}\otimes \mathbb{I}_B\nonumber\\
    &=\mathbb{I}_{\widetilde{A}}\otimes \ket{\Psi_\mu}\bra{\Psi_0}_{AA'}(\mathbb{I}_{A}\otimes \sigma_\mu^{(A')})\otimes \mathbb{I}_B.
\end{align}
Let
\begin{align}
    \ket{\Psi}_{\widetilde{A}A}=\sum_{i=e,g}\sqrt{p_i}\ket{\psi_i}_{\widetilde{A}}\ket{\phi_i'}_{A}
\end{align}
be the Schmidt decomposition, where $\{\ket{\psi_i}_{\widetilde{A}}\}_{i=e,g}$ and $\{\ket{\phi_i'}_{A}\}_{i=e,g}$ are orthonormal basis.
The state is expanded as
\begin{widetext}
\begin{align}
    \xi_{\widetilde{A}B}(\mu)&=\sum_{i,j=g,e}\sqrt{p_ip_j}\ket{\psi_i}\bra{\psi_j}_{\widetilde{A}}\otimes \left(\bra{\Psi_0}_{AA'}\mathbb{I}_{A}\otimes \sigma_\mu^{(A')}\otimes \mathbb{I}_B(\ket{\phi_i'}\bra{\phi_j'}_{A}\otimes  \rho_{A'B})\mathbb{I}_{A}\otimes \sigma_\mu^{(A')}\otimes \mathbb{I}_B\ket{\Psi_0}_{AA'}\right)
\end{align}
After the measurement, Bob performs a unitary operation $u_\mu$. The unnormalized state is given by
\begin{align}
    \mathbb{I}_{\widetilde{A}}\otimes u_{\mu}^{(B)}\xi_{\widetilde{A}B}(\mu)\mathbb{I}_{\widetilde{A}}\otimes u_{\mu}^{(B)}&=\sum_{i,j=g,e}\sqrt{p_ip_j}\ket{\psi_i}\bra{\psi_j}_{\widetilde{A}}\otimes \left(\bra{\Psi_0}_{AA'}\ket{\phi_i}\bra{\phi_j}_{A}\otimes( \sigma_\mu^{(A')}\otimes u_\mu^{(B)}  \rho_{A'B}\otimes \sigma_\mu^{(A')}\otimes u_\mu^{(B)})\ket{\Psi_0}_{AA'}\right)
\end{align}
Summing over $\mu$, we get
\begin{align}
    \xi_{\widetilde{A}B}&\coloneqq\sum_{\mu=0}^3\mathbb{I}_{\widetilde{A}}\otimes u_{\mu}^{(B)}\xi_{\widetilde{A}B}(\mu)\mathbb{I}_{\widetilde{A}}\otimes u_{\mu}^{(B)}\nonumber\\
    &=\sum_{i,j=g,e}\sqrt{p_ip_j}\ket{\psi_i}\bra{\psi_j}_{\widetilde{A}}\otimes \left(\bra{\Psi_0}_{AA'}\ket{\phi_i'}\bra{\phi_j'}_{A}\otimes\left(\sum_{\mu=0}^3 \sigma_\mu^{(A')}\otimes u_\mu^{(B)}  \rho_{A'B}\otimes \sigma_\mu^{(A')}\otimes u_\mu^{(B)}\right)\ket{\Psi_0}_{AA'}\right).
\end{align}
\end{widetext}

The entanglement resource state is given by
\begin{align}
    \rho_{A'B}=
    \begin{pmatrix}
    1-\mathcal{L}_{A'A'}-\mathcal{L}_{BB}&0&0&\mathcal{M}^*\\
    0&\mathcal{L}_{BB}&\mathcal{L}_{A'B}&0\\
    0&\mathcal{L}_{BA'}&\mathcal{L}_{A'A'}&0\\
    \mathcal{M}&0&0&0
    \end{pmatrix}+\mathcal{O}(\lambda^4)
\end{align}
in a basis $\{\ket{gg}_{A'B},\ket{ge}_{A'B},\ket{eg}_{A'B},\ket{ee}_{A'B}\}$. Here, we adopt $u_\mu=\sigma_\mu v_B$ 
\begin{align}
     u_\mu=\sigma_\mu v_B
\end{align}
as Bob's local unitary operation, where
\begin{align}
    v_B\coloneqq e^{-\ii\varphi}\ket{e}\bra{e}_B+\ket{g}\bra{g}_B
\end{align}
and $\varphi\in\mathbb{R}$ is the argument of $\mathcal{M}$, i.e., $\mathcal{M}=|\mathcal{M}|e^{\ii \varphi}$. The operation $v_B$ eliminates the phase in $\mathcal{M}$, i.e.,
\begin{align}
    &\mathbb{I}_{A'}\otimes v_B\rho_{A'B}\mathbb{I}_{A'}\otimes v_B^\dag\nonumber\\
    &=
    \begin{pmatrix}
    1-\mathcal{L}_{A'A'}-\mathcal{L}_{BB}&0&0&|\mathcal{M}|\\
    0&\mathcal{L}_{BB}&\widetilde{\mathcal{L}_{A'B}}&0\\
    0&\widetilde{\mathcal{L}_{A'B}}^*&\mathcal{L}_{A'A'}&0\\
    |\mathcal{M}|&0&0&0
    \end{pmatrix}+\mathcal{O}(\lambda^4),
\end{align}
where $\widetilde{\mathcal{L}_{A'B}}\coloneqq \mathcal{L}_{A'B}e^{\ii\varphi} $. Since
\begin{align}
    &\frac{1}{4}\sum_{\mu=0}^3\sigma_{\mu}^{(A')}\otimes u_\mu\rho_{A'B}\sigma_{\mu}^{(A')}\otimes u_\mu
    =\eta_{A'B}
\end{align}
holds for $\eta_{A'B}$ defined in Eq.~\eqref{eq:density_mat_noisy_qst}, we get Eq.~\eqref{eq:xi}.

\section{On the role of $v_B$}\label{sec:phase_cancelling_op}
We here analyze the case where Bob implements Step(4) instead of (4'). In other words, he performs $\sigma_\mu$ instead of $\sigma_\mu v_B$ when he receives the measurement outcome $\mu$ from Alice. In this case, $\eta_{A'B}$ in Eq.~\eqref{eq:density_mat_noisy_qst} changes into
\begin{align}
    \eta_{A'B}&= \frac{1}{4}\sum_{\mu=0}^3\sigma_\mu^{(A')}\otimes \sigma_\mu^{(B)}\rho_{A'B}\sigma_\mu^{(A')}\otimes \sigma_\mu^{(B)}\\
    &=
    \begin{pmatrix}
        \frac{1}{2}-\mathcal{L}&0&0&\mathrm{Re}\left(\mathcal{M}\right)\\
        0&\mathcal{L}&\mathrm{Re}\left(\mathcal{L}_{A'B}\right)&0\\
        0&\mathrm{Re}\left(\mathcal{L}_{A'B}\right)&\mathcal{L}&0\\
        \mathrm{Re}\left(\mathcal{M}\right)&0&0&\frac{1}{2}-\mathcal{L}
    \end{pmatrix}+O(\lambda^4).
\end{align}
Similar to the result in the main text, we can explicitly evaluate the teleported negativity if $\{\braket{\phi_i|\phi_k'}\}_{i,k}=\delta_{ik}$. Instead of Eqs.\eqref{eq:negativity_teleportation_general}, \eqref{eq:negativity_teleportation_general_2nd} and \eqref{eq:negative_ev}, we have
\begin{align}
    \mathcal{N}\left(\xi_{\widetilde{A}B}\right)&=\mathcal{N}^{(2)}\left(\xi_{\widetilde{A}B}\right)+\mathcal{O}(\lambda^3),\\
    \mathcal{N}^{(2)}\left(\xi_{\widetilde{A}B}\right)&\coloneqq\max\{0,-E''\},
\end{align}
where
\begin{align}
    E''&\coloneqq \mathcal{L}-\sqrt{\mathcal{L}^2(1-4p(1-p))+4p(1-p)\left(\mathrm{Re}\left(\mathcal{M}\right)\right)^2}.
\end{align}
Since $-E''\leq -E'$, the teleported negativity decreases if Bob performs $\sigma_\mu$ instead of $\sigma_\mu v_B$. In other words, the phase-eliminating operation $v_B$ makes it possible to transfer negativity more efficiently. 

If we further assume that $\widetilde{A}$ and $A$ are initially maximally entangled, i.e., $p=1/2$, we get
\begin{align}
    \mathcal{N}^{(2)}\left(\xi_{\widetilde{A}B}\right)=\max\{0,|\mathrm{Re}\left(\mathcal{M}\right)|-\mathcal{L}_{A'A'}\}. 
\end{align}
Therefore, unless $|\mathrm{Re}\left(\mathcal{M}\right)|=|\mathcal{M}|$, it holds
\begin{align}
    \mathcal{N}^{(2)}\left(\xi_{\widetilde{A}B}\right)<\mathcal{N}^{(2)}\left(\rho_{A'B}\right).
\end{align}
This fact highlights the difference from Eq.~\eqref{eq:optimality}, showing that the teleportation protocol is not optimal if Bob performs $\sigma_\mu$ instead of $\sigma_
\mu v_B$. 

\end{document}